\documentclass[12pt]{article}

\usepackage{latexsym}

\usepackage{eufrak}

\usepackage[mathcal]{euscript}

\skip\footins 12pt                   
\newcommand{\be}{\vspace*{6pt} \begin{equation}}
\newcommand{\ee}{\vspace*{10pt} \end{equation}}

\renewcommand{\section}[1]{\addtocounter{section}{1}
                           \setcounter{subsection}{0}
                           \setcounter{subsubsection}{0}
                           \vspace{20pt}
                           \begin{center}
                           {\large \thesection \ #1}
                           \end{center}
                           \vspace{20pt}}

\renewcommand{\subsection}[1]{\addtocounter{subsection}{1}
                           \setcounter{subsubsection}{0}
                           \vspace{20pt}
                           \begin{center}
                           \thesubsection \ #1
                           \end{center}
                           \vspace{20pt}}
\renewcommand{\subsubsection}[1]{\addtocounter{subsubsection}{1}
                           \vspace{20pt}
                           \begin{center}
                           \thesubsubsection \ #1
                           \end{center}
                           \vspace{20pt}}

\newcommand{\ie}{{\it i.e.}}         

\newcommand{\eg}{{\it e.g.}}

\newcommand{\ident}{1\!{\rm I}}                                 
\newcommand{\cH}{\cal{H}}                                       
\newcommand{\cB}{\cal{B}}                                       
\newcommand{\ga}{\alpha}                                        
\newcommand{\gb}{\beta}                                         
\newcommand{\gab}{\alpha\beta}                                  
\newcommand{\gn}{\nu}                                           
\newcommand{\gw}{\omega}                                        

\renewcommand{\epsilon}{\varepsilon}
\renewcommand{\Im}{\mbox{{\sf Im}}}                             
\newcommand{\Tr}[1]{\mbox{{\sf Tr}$[#1]$}}                      
\newcommand{\abs}[1]{\left|#1\right|}

\newcommand{\bra}[1]{\langle #1\vert}                           
\newcommand{\ket}[1]{\vert #1\rangle}                           
\newcommand{\braket}[2]{\langle #1\vert #2\rangle}              

\begin{document}

\baselineskip 12pt

\thispagestyle{empty}
\begin{center}

\vspace*{20pt}

\noindent {\large  {\bf   Dynamics for Density Operator 
Interpretations of Quantum Theory}}

\vspace{40pt}
\noindent Guido Bacciagaluppi

\noindent Balliol College, Oxford

\noindent and

\noindent Sub-Faculty of Philosophy

\noindent University of Oxford

\vspace{20pt}
\noindent Michael Dickson

\noindent Department of History and Philosophy of Science

\noindent Indiana University

\vspace{20pt}
\noindent {\bf Abstract}
\end{center}

\begin{quotation}
\noindent We first introduce and discuss density operator 
interpretations of quantum theory as a special case of a more general 
class of interpretations, giving special attention to a version that 
we call the `atomic version'.  We then review some crucial parts of 
the theory of stochastic processes (the proper context in which to 
discuss dynamics), and develop a general framework for specifying a 
dynamics for density operator interpretations.  This framework admits 
infinitely many empirically equivalent dynamics.  We give some 
examples, and discuss some of the properties of one of them.
\end{quotation}

\newpage

\setcounter{page}{1}
\thispagestyle{empty}

\begin{center}

\vspace{12pt}

\noindent Dynamics for Density Operator Interpretations of Quantum 
Theory

\vspace{20pt}
\noindent {\bf Abstract}
\end{center}

\begin{quotation}
\noindent We first introduce and discuss density operator 
interpretations of quantum theory as a special case of a more general 
class of interpretations, giving special attention to a version that 
we call the `atomic version'.  We then review some crucial parts of 
the theory of stochastic processes (the proper context in which to 
discuss dynamics), and develop a general framework for specifying a 
dynamics for density operator interpretations.  This framework admits 
infinitely many empirically equivalent dynamics.  We give some 
examples, and discuss some of the properties of one of them.
\end{quotation}

\vspace{20pt}

\section{Density Operator Interpretations as Modal Interpretations}

\vspace{-30pt}

\subsection{Introduction to Modal Interpretations}

\noindent One way to `interpret' quantum mechanics is to say, for any
system, what properties are possessed, or what observables have a
definite value.  The theorem of Kochen and Specker (1967) shows that,
under certain constraints (which we will take for granted), one
cannot take {\it every} observable to have a definite value.  Hence
the central interpretive question: Which observables have a definite
value?

There are in the literature a number of proposed answers to this 
question,$^1$ the aim of which is to prescribe, at each time, a set of 
possible properties (or, equivalently, a set of definite-valued 
observables), \ie, those properties that may be possessed by a system 
(as opposed to the properties, familiar from standard quantum 
mechanics, that are neither possessed nor not possessed).  In 
addition, of course, one needs a probability measure over 
the possible properties.  We are thus given an answer to any question 
of the form: At time $t$, what properties may be possessed, and for 
any such property, what is the probability that it is possessed?

These `properties that may be possessed', the possible properties of a 
system at a time, are probably best required to form either an algebra 
or a partial algebra under the quantum-algebraic (lattice-theoretic) 
operations of meet, join, and orthocomplement.  After all, probability 
measures are generally taken to be defined over {\it algebras} (or 
partial algebras) of events.$^2$ As it happens, many proposals choose 
an algebra of a certain kind, dubbed a faux-Boolean algebra.  These 
algebras are constructed (in Hilbert space) as follows.  Choose any 
set, $S$, of mutually orthogonal subspaces in the Hilbert space.  Let 
$S^{\perp}$ be the set of all subspaces orthogonal to the span of $S$.  
The algebraic closure (under meet, join, and orthocomplement) of $S 
\cup S^{\perp}$ is a faux-Boolean algebra.  We will mainly be 
concerned with proposals that choose for their algebra of possible 
properties a faux-Boolean algebra, though we will point out when our 
results generalize to other proposals.

One must, of course, adopt the quantum-mechanical probability measure 
over the algebra of possible properties.  That is, for any system with 
the state (statistical operator), $W$, the probability that the system 
possesses the property $P$ is given by $\Tr{WP}$, where $P$ is, of 
course, an element in the system's faux-Boolean algebra of possible 
properties.  (We do not distinguish notationally among projection 
operators, subspaces, and properties.) Moreover, all extant proposals 
of the sort considered here (\ie that choose a faux-Boolean algebra) 
choose $S$ so that every property in $S^{\perp}$ has probability zero.  
That is, all of these proposals make the faux-Boolean algebra of 
properties state-dependent, and do so in such a way that every element 
of $S^{\perp}$ is in the null space of $W$.  (This requirement 
guarantees that the probability theory generated by modal 
interpretations is effectively classical.  See Dickson (1995a, 1995b) 
for discussion.  See also Bell and Clifton (1995) and Zimba and 
Clifton (1998) for discussions of the algebraic structure of possible 
properties.)

For every property, $P$, in a faux-Boolean algebra, $\cB$, of possible 
properties, then, we may say that $P$ is either possessed or not by a 
system (at a given time).  The complete set of possessed properties 
for a given system (at a given time) is therefore a map, $m:{\cB} 
\rightarrow\{0,1\}$ (`0' for `not possessed' and `1' for `possessed').  
It is easy to show that, because we have adopted the 
quantum-mechanical probability measure over $\cB$, $m$ must be an 
ultrafilter on $\cB$.  That is, $m$ must map one and only one atom, 
$P_{\mbox{{\scriptsize atom}}}$, of $\cB$ to 1, and for every other 
element, $P$, of $\cB$, $m(P) = 1$ if and only if $P 
P_{\mbox{{\scriptsize atom}}} = P_{\mbox{{\scriptsize atom}}}$ (or, in 
terms of subspaces, $P_{\mbox{{\scriptsize atom}}} \subseteq P$).  One 
can therefore specify the {\it complete} state of a system by 
specifying an atom in its faux-Boolean algebra.  Moreover, as we 
mentioned, with probability 1 the complete state of a system will be 
specified by an element of the set, $S$, used in the construction of 
its faux-Boolean algebra.  Henceforth, therefore, we will mostly 
restrict our attention to the elements of $S$, and we refer to the 
actually possessed element of $S$ as `the complete state' of a system.

\subsection{The Atomic Version}

We will concentrate our attention on a proposal that we
call the `atomic version', but we begin with the version due to Dieks
(1988, 1989) and its generalization by Vermaas and Dieks (1995).
(These are related to versions due to Kochen (1985) and Healey (1989).
See note 1 for further references.) In the above terminology, Vermaas and
Dieks construct a set, $S(t)$, at each
time, $t$, as follows: for any system with a quantum-mechanical state,
$W(t)$, the elements of $S(t)$ are the elements that correspond to non-zero
eigenvalues in the (unique) spectral resolution of $W(t)$.

The apparent simplicity of this proposal is deceiving.  The
Vermaas--Dieks proposal is meant to apply to {\it every} system
individually.  We should therefore ask: What {\it counts} as a
`system'?  Possibly, Dieks' original intuition was that every Hilbert space,
${\cH}^{\gn}$ that appears in {\it some} factorization,
$\cH_{\mbox{{\scriptsize univ}}} = \cH^{\ga} \otimes
\cH^{\gb} \otimes \ldots \cH^{\gw}$ of the Hilbert space,
$\cH_{\mbox{{\scriptsize univ}}}$, of the universe corresponds to a 
system, in the above sense.  (The index $\gn$ ranges over the Greek 
superscripts $\ga, \gb, \ldots$.) One then uses the statistical 
operator for each such subsystem (found by tracing out the rest of the 
universe), to establish a faux-Boolean algebra for that subsystem.  
However, it has been shown (Bacciagaluppi 1995) that this option leads 
to a Kochen--Specker contradiction.  Apparently, the only way around 
this contradiction is to adopt some form of contextuality.  (See 
Bacciagaluppi and Vermaas (1997) for further discussion.) Dieks (1997) 
has, in fact, chosen to follow Healey in supposing that there is a 
{\it preferred factorization} of the Hilbert space of the universe 
(though for Dieks, the preferred factorization should be chosen, in a 
sense pragmatically, to depend on the relevant interactions).  
Subsystems of the universe then correspond to the factor spaces that 
appear in the preferred factorization of $\cH_{\mbox{{\scriptsize 
univ}}}$.

As an aside, we note that the idea of a preferred factorization is
not, perhaps, as {\it ad hoc} as it might at first appear.  After all,
assuming that the universe is really made up of, say, electrons,
quarks, and so on, it makes good sense to take these objects to be the
`real' constituents of the universe, \ie, the bearers of properties
that do not necessarily supervene on the properties of subsystems.
Indeed, it would appear strange, if not downright silly, to
suppose that, for example, a `system' composed of the spatial degrees
of freedom of some electron and the spin degrees of freedom of some
atom is a genuine subsystem of the universe, deserving of its own
properties (apart from those properties that it inherits by virtue of
its being composed of two other systems).

However, even this proposal runs into difficulties.  Consider, for
example, a system whose Hilbert space is $\cH^{\gab} = \cH^{\ga}
\otimes \cH^{\gb}$, and one of its subsystems, whose Hilbert space is
$\cH^{\ga}$.  (We refer to such systems and subsystems by the
superscripts that label the Hilbert spaces, $\ga, \gb, \gab$.) It can
easily happen that a spectral projection, $P^{\gab}$, of $W^{\gab}$
(the quantum-mechanical state of $\gab$) does not commute with
$P^{\ga} \otimes \ident^{\gb}$, where $P^{\ga}$ is a spectral projection of
$W^{\ga}$ and $\ident^{\gb}$ is the identity on $\cH^{\gb}$.  But then it is
not obvious that we can escape a Kochen--Specker contradiction (and, indeed,
Clifton (1996) has derived one), or even that we can define a joint
probability for $\gab$ to possess
$P^{\gab}$ and $\ga$ to possess $P^{\ga}$. In fact, it is well known that
there is, in general, no expression for the joint probability for
non-commuting projections that is valid in every quantum-mechanical
state.  (Whether or not we wish to say that $\gab$ possesses $P^{\ga}
\otimes \ident^{\gb}$ whenever $\ga$ possesses $P^{\ga}$---an assumption
necessary for deriving a Kochen--Specker contradiction---a joint
probability for $\gab$ to possess $P^{\gab}$ and $\ga$ to possess
$P^{\ga}$ will induce a joint probability measure for $P^{\gab}$ and
$P^{\ga} \otimes \ident^{\gb}$.)

On the other hand, we do not need a general expression for the joint 
probability of any two non-commuting projections, but only expressions 
that are valid for limited sets of non-commuting projections.  
Nevertheless, while some have tried to find such expressions, and have 
succeeded in special cases (Vermaas 1996), no generally acceptable 
expression has yet been found.  Indeed, recent results (Vermaas 1997) 
suggest that no satisfactory expression will be found.

Because of these problems, we propose to adopt a still more 
conservative approach, which we call the `atomic version'.  This 
version adopts a preferred factorization of the Hilbert space for the 
universe, and assigns faux-Boolean algebras of possible properties 
directly only to those subsystems corresponding to atomic factors in 
the preferred factorization (\ie, those that are not themselves tensor 
products of factors appearing in the preferred factorization).  All 
other subsystems inherit properties from these `atomic' subsystems by 
the principle of property composition: if two subsystems, $\ga$ and 
$\gb$, possess the properties $P^{\ga}_{m}$ and $P^{\gb}_{n}$, 
respectively, then the system composed of $\ga$ and $\gb$ possesses 
the property $P^{\ga}_{m} \otimes P^{\gb}_{n}$.

For example, suppose that the Hilbert space for the universe has the
preferred factorization $\cH_{\mbox{{\scriptsize univ}}} = \cH^{\ga}
\otimes \cH^{\gb} \otimes \ldots \otimes \cH^{\gw}$. The atomic version
assigns a faux-Boolean algebra to each of the subsystems $\ga , \ldots 
, \gw$.  Following Vermaas and Dieks (1995), it does so by letting 
$S^{\gn}$ (the set used to generate the faux-Boolean algebra, 
${\cB}^{\gn}$, for the atomic system, $\gn$) be the set of projections 
with non-zero eigenvalues in the (unique) spectral resolution of 
$W^{\gn}$, where $W^{\gn}$ is the quantum-mechanical state of the 
atomic system, $\gn$.

As we said, in the atomic version, all compound systems have only
those properties that they inherit from their constituent atomic
subsystems.  Therefore, the algebra of properties for a compound
system is just the Cartesian product of the algebras of properties for
the constituent atomic subsystems.  To complete the atomic version,
therefore, we need only the joint probabilities for properties of all
subsystems.  The other joint probabities are obtained by additivity.  As
Vermaas and Dieks (1995) have already noted, the obvious candidate is
\be
\Pr \left( { P^{\ga}_{i_\ga}, P^{\gb}_{i_\gb}, \ldots , P^{\gw}_{i_\gw}}
\right) =
\Tr{WP^{\ga}_{i_\ga} \otimes P^{\gb}_{i_\gb} \otimes \ldots \otimes
P^{\gw}_{i_\gw}}.
\ee

From this expression, it is clear that while atomic systems $\ga$ and
$\gb$ might have non-zero probability to possess the properties
$P^{\ga}$ and $P^{\gb}$, respectively, it might still happen that the
compound system $\gab$ has zero probability to possess the property
$P^{\ga}\otimes P^{\gb}$.  (Think, for example, of the singlet state
of two spin-$1/2$ particles.) In this case one might like to say that
$P^{\ga}\otimes P^{\gb}$ should {\it not} be in the set, $S^{\gab}$,
used to generate the faux-Boolean algebra of properties for the
compound system $\gab$.  However, such a policy would make the problem
of finding a dynamics even harder than it already is.  The reason is
that finding a dynamics is made more difficult when the cardinality of
the set, $S^{\gab}$, can change in time.  (This point should become
clear in Sections 2 and 3.) Hence we adopt the policy that the
faux-Boolean algebra of properties for a compound system is indeed
generated by the set
\be
\label{eq:composite}
S := \left\{ { P = P^{\ga}_{i_{\ga}} \otimes \ldots \otimes
P^{\gw}_{i_{\gw}}\ |\ P^{\ga}_{i_{\ga}} \in S^{\ga}, \ldots ,
P^{\ga}_{i_{\gw}} \in S^{\gw} } \right\},
\ee
where the Hilbert space for the compound system is $\cH = \cH^{\ga}
\otimes \ldots \otimes \cH^{\gw}$, $\ga, \ldots , \gw$ are atomic systems,
and for all $\gn$, $S^{\gn}$ is the set used to generate the faux-Boolean
algebra of properties for the atomic system $\gn$.  This construction
differs from the perhaps more natural one, given by the additional
constraint that $\Tr{WP} \neq 0$ (where $W$ is the state of the
compound system), only on properties that have zero probabilty.

To finish this discussion, and to foreshadow the discussion of Section 
3, note that in general the sets $S^{\gn}$ are time-dependent.  Our 
work here may therefore be seen as a generalization of work by Bohm 
(1952), Bell (1987), Vink (1993), and Bub (1996, 1997).  They consider 
a dynamics for a `preferred observable', $R$, considered to have a 
definite value at all times.  For us, the preferred observable is 
time-dependent.  That is, we may take it to be some observable, 
$R(t)$, whose eigenprojections are the elements of $S(t)$ (the set $S$ 
as given by (\ref{eq:composite}) at some time, $t$), and whose 
eigenvalues are arbitrary.  Indeed, as we shall see, our work applies 
to any time-dependent preferred observable whose eigenprojections and 
eigenvalues evolve differentiably.  (We return to this point in 
greater detail later.)

There is plenty more one might say about the atomic version, but we
will not linger over its motivation or consequences here.  As we said
above, our results will be geared mainly towards the atomic version,
though we will note how and under what conditions our results can be
applied to non-atomic versions.

(Lest we appear overly optimistic about the prospects for the atomic
version, we note that one outstanding problem has not been solved: we have
not shown that it will attribute `the right' properties to macroscopic
objects. For example, we have not shown that it entails that macroscopic
objects are well-localized. This question has been addressed in detail,
with ultimately negative results, for non-atomic interpretations.$^{3}$)

\subsection{The Problem}

Thus far we have specified, for every time $t$, the set of possible 
modal states, $S(t)$, and a probability measure over this set.  
However, one wants to know more.  We take it to be crucial to answer 
dynamical questions, of the form: Given that a system possesses 
property $P$ at time $s$, what is the probability that it will possess 
property $P'$ at time $t$ ($t > s$)?  In other words, we need a {\it 
dynamics} of possessed properties.

Some may consider a dynamics of possessed properties to be
superfluous.  After all, could quantum mechanics not get away with
just single-time probabilities?  Why can we not settle for an
interpretation that supplements standard quantum mechanics {\it only}
by providing in a systematic way a set (the set of possible
properties) over which single-time probabilities are defined?  If
we require of this set that it include the everyday properties of
macroscopic objects,$^4$ then what more do we need?

What we need is an assurance that the {\it trajectories} of possessed 
properties are, at least for macroscopic objects, more or less as we 
see them to be.  For example, we should require not only that the book 
at rest on the desk has a definite location, but also that, if 
undisturbed, its location relative to the desk does not change in 
time.  Hence one cannot get away with simply specifying the definite 
properties at each time.  We need also to be shown that this 
specification is at least {\it compatible} with a reasonable dynamics.  
Even better, we would like to see the dynamics explicitly.$^5$ (As we 
will note below, it is trivial to define an `unreasonable' dynamics, 
namely, one in which there is no correlation from one time to the 
next.  In such a case, the book on the table might {\it not} remain at 
rest relative to the table, even if undisturbed.  We take it that such 
dynamics are not very interesting, and fail to provide any assurance 
that we can describe the world more or less as we think it is.$^6$)

Our main task in this paper is to show how to construct a dynamics 
describing time-evolution of the complete state of a system.  We will 
investigate some features of one example, with an eye towards 
suggesting that there are likely to be at least some dynamics that are 
`reasonable'.

The structure of the rest of this paper is as follows.  In the next 
section, we will discuss the general framework of stochastic 
processes, the appropriate mathematical theory for the rigorous 
description of a dynamics.  In particular, we will discuss the 
question of how to find finite-time transition probabilities.  This 
problem is much-discussed among theorists of stochastic processes.  In 
Section 3, we develop a general framework for the description of 
dynamics for complete states.  Here we show how to specify 
infinitesimal parameters that are guaranteed to return the single-time 
probabilities already defined via quantum theory.  In Section 4, we 
discuss some constraints on dynamics first derived by Vermaas (1996), 
and we give some examples of dynamics, spending particular attention 
on one that we call the generalized Schr\"odinger dynamics.  In 
Section 5 we discuss some of the properties of the generalized 
Schr\"odinger dynamics, and we mention some problems for future 
research.  (A review of research on dynamics will appear in 
Bacciagaluppi (1998a).)

\section{Preparatory Discussion of Dynamics}

\vspace{-30pt}

\subsection{Stochastic Processes}

It is evident that the dynamics we are after will be genuinely
probabilistic.  This point can be seen in a trivial example.  Let
\be
\ket{\Psi(t)}:=\sum_{i} c_{i}(t)\ket{\ga_{i}}\otimes \ket{\gb_{i}},
\ee
\noindent where $\ket{\ga_{i}} \in \cH^{\ga}$ and $\ket{\gb_{i}} \in
\cH^{\gb}$.  Then the spectral resolution of $W^{\ga}(t)$ is at all
times given by $\{P^{\ga}_{i}\}\ (= \{\ket{\ga_{i}}\bra{\ga_{i}}\})$,
unless there happens to be a degeneracy (in which case the projections
in the unique spectral resolution of $W^{\ga}$ would be given by sums
of elements of $\{P^{\ga}_{i}\}$)---we will assume for this example
that there are no degeneracies.  But the probabilities attached to
these spectral projections will change, due to the time-dependence of
the $c_{i}(t)$.  Now, consider an ensemble of systems, all with
state vector $\ket{\Psi(t)}$ and with modal states distributed across
$\{P^{\ga}_{i}\}$ according to the quantum-mechanical probability
measure, $|c_{i}(t)|^{2}$.  As the distribution, $|c_{i}(t)|^{2}$,
changes in time, some members of the ensemble must make transitions
among the $\{P^{\ga}_{i}\}$ in order to preserve the distribution.

Why can these transitions not be deterministic?  Because, assuming 
that the compete state specifies completely the physical state of the 
system, there can be nothing to distinguish those systems that make a 
transition from, say, $P^{\ga}_{1}$ to, say, $P^{\ga}_{2}$ from those 
systems that make a transition from $P^{\ga}_{1}$ to some property 
other than $P^{\ga}_{2}$ (or make no transition at all).  There is 
therefore nothing in the theory that could have `determined' the 
transition from $P^{\ga}_{1}$ to $P^{\ga}_{2}$.

Deterministic `equations of motion' for the complete states are 
therefore out of the question.  Instead, we need stochastic equations 
of motion.  The most general and powerful framework within which to 
find and study such equations is the theory of stochastic processes.  
We shall therefore put our question into this framework, allowing us 
to borrow and adapt some well-known results from that theory.

We remind the reader of some basic definitions.  (There are many books 
on the topic.  Classic texts are Doob (1953) and, in particular, 
Feller (1968).) A {\bf random variable}, $V$, is a map from a sample 
space (the set of all atomic events in some algebra of events) to the 
real numbers.  The probability measure over (the algebra generated by) 
the sample space induces a probability measure over (the algebra 
generated by) the range of $V$.  We therefore ignore the sample space 
itself, and consider directly the probability measure over the range 
of $V$, also called the {\bf state space} of $V$, writing:
\be
p^{V}_{i} := p(V = i),
\ee
\noindent defined for all $i$ in the state space of $V$.  Because the
context makes the meaning clear, we will usually omit the
superscripted $V$, writing $p_{i}$.

A {\bf stochastic process}, $V_{t}$ is an indexed family of random
variables with a common state space.  For us, the index $t$ will be
continuous, and it will represent time.  Moreover, for us, the state
space will be a set of integers, $I$.  Single-time probability
distributions are written:
\be
p^{V_t}_{i}(t) := p(V_{t} = i)
\ee
\noindent for all $i$ in the state space, $I$, of $V_{t}$.  Again, we
usually write $p_{i}(t)$.

Given a complete set of finite-time transition probabilities, that
is, the joint distribution functions,
\be
p(V_{t_{1}} = i_{1}, V_{t_{2}} = i_{2}, \ldots , V_{t_{n}} = i_{n}),
\ee
for every set of times, $\{t_{1}, \ldots t_{n}\}$ (where of course
$i_{m} \in I$ for all $m$ in $\{1, \ldots , n\}$), we can define the
finite-time transition probabilities in the usual way by
\be
\label{eq:finitetimetrans}
p(V_{t} = i | V_{t_{1}} = i_{1}, \ldots , V_{t_{n}} = i_{n}) :=
\frac{p(V_{t_{1}} = i_{1}, \ldots , V_{t_{n}} = i_{n}, V_{t} =
i)}{p(V_{t_{1}} = i_{1}, \ldots , V_{t_{n}} = i_{n})}\, ,
\ee
where $t > t_m$ for all $m$ in $\{1, \ldots , n \}$.  Equation
(\ref{eq:finitetimetrans}) represents the most general form for the
finite-time transition probabilities, and without making any further
restrictions, a complete dynamics requires a complete set of such
probabilities.  However, we will restrict attention here to Markov
processes, that is, processes that obey the Markov property:
\begin{quotation}
{\bf Markov Property}:  Whenever $t_{1} < t_{2} < \ldots < t_{n} < t$,
\be
p(V_{t} = i | V_{t_{1}} = i_{1}, \ldots V_{t_{n}} = i_n) =
p(V_{t} = i | V_{t_{n}} = i_n).
\ee
\end{quotation}
\noindent We therefore need to find only a complete set of {\bf transition
probability functions},
\be
p_{ji}(t,s) := p(V_{t} = j | V_{s} = i),
\ee
for all $s \leq t$.  We require, for obvious reasons, that
$p_{ii}(t,t) = 1$ and $p_{ji}(t,t) = 0$ for $i \neq j$.  (Note
that the transition probability function is read from right to left:
$p_{ji}(t,s)$ is the probability of a transition {\it from} $i$ at time $s$
{\it to} $j$ at time $t$.)

Now, one could try to specify directly all of the $p_{ji}(t,s)$,
but this strategy is too unwieldy.  It is more common, and more
convenient in the end, to begin by defining the so-called
infinitesimal parameters (defined in the next subsection); then one uses
the infinitesimal parameters to build up the finite-time transition
probabilities.

Of course, there is one trivial way to specify directly all of the
$p_{ji}(t,s)$.  It is the case of no correlation over time.  That is,
for all $s,t,i,j$, define
\be
\label{eq:trivial}
p_{ji}(t,s) := p_{j}(t).
\ee
Such transition probabilities describe the `unreasonable' dynamics
that we mentioned above.  They obviously return the correct
single-time probabilities.

\subsection{The Infinitesimal Parameters}

As we said, we are seeking a dynamics that is more reasonable than
(\ref{eq:trivial}), and the best way to do so is to begin with the
infinitesimal parameters, defined below.  However, this procedure
brings with it some complications that require further discussion.  In
this section, we first give some definitions, and then discuss the
problems that one encounters when trying to construct finite-time
transition probabilities from the infinitesimal parameters.  This
problem is a standard one in the theory of Markov processes.  Our
discussion will therefore be brief, and proofs will be omitted.  (One
can find discussions of the results that we quote in standard books on
stochastic processes.  See, for example, Doob (1953), Feller (1968), 
and Gikhman and Skorokhod (1974--1979).)

The {\bf infinitesimal parameters}, of a stochastic process (not
necessarily Markovian) are defined, for $j\neq i$, as the quantities
\be
\label{eq:inf1}
t_{ji}(t):=\lim_{\epsilon\rightarrow 0}
\frac{p_{ji}(t+\epsilon,t) - p_{ji}(t,t)}{\epsilon} =
\lim_{\epsilon\rightarrow 0}
\frac{p_{ji}(t+\epsilon,t)}{\epsilon},
\ee
if the limit on the right-hand side exists in $[0,\infty]$.  (We have
used $p_{ji}(t,t) = 0$ for $j \neq i$.) For $j=i$ they are
\be
\label{eq:inf2}
t_{ii}(t):= \lim_{\epsilon\rightarrow 0}\frac{p_{ii}(t+\epsilon,t) -
p_{ii}(t,t)}{\epsilon} = \lim_{\epsilon\rightarrow
0}\frac{p_{ii}(t+\epsilon,t)-1}{\epsilon},
\ee
when the limit exists in $[-\infty,0]$.  (We have used $p_{ii}(t,t) =
1$.)

>From this definition of $t_{ji}$, it follows that
\be
\sum_j t_{ji}(t)=0.
\label{eq:normalisation}
\ee
Intuitively, $t_{ji}(t)$ (for $j\neq i$), which is always positive, is
the rate at which $j$ gains probability at the expense of $i$, while
$t_{ii}(t)$ (which is always negative) is the rate at which $i$ loses
probability.  Equation (\ref{eq:normalisation}) expresses the
conservation of probability.

There is an alternative motivation of the infinitesimal parameters.
Define $t_{i}(t)dt$ to be the probability that the process makes a
transition from state $i$ during the interval $[t,t+dt]$, assuming it
is in the state $i$ at time $t$.  Define $\Pi_{ji}(t)$ to be the
probability of a jump from $i$ to $j$ during the same time interval,
given that the process is in the state $i$ at time $t$ and that a
transition takes place.  It follows directly from the definitions
above (when the $t_{ji}$ are finite) that
\be
\label{eq:ti(t)}
t_i(t) = -t_{ii}(t)
\ee
and
\be
\label{eq:Piji(t)}
\Pi_{ji}(t) = \frac{t_{ji}(t)}{t_{i}(t)}.
\ee
(Note that $\sum_j\Pi_{ji}(t)=1$.)

Equation (\ref{eq:ti(t)}) establishes a connection between the
infinitesimal parameters and the waiting time in state $i$.  In
particular, one can show that the waiting time in state $i$ after
having arrived there at time $s$ is exponentially distributed with
parameter $t_{i}(u)$.  In other words, the probability that a system
that arrived in the state $i$ at time $s$ will remain there at least
until time $t$ is
\be
\label{eq:waittime}
p_{ii}(t,s) = \exp \left[ { -\int_{s}^{t}t_{i}(u)du } \right].
\ee

Using (\ref{eq:ti(t)})--(\ref{eq:waittime}), one can already see a
connection between the infinitesimal parameters and the finite-time
transition probabilities, by using $t_{i}(t)$ and $\Pi_{ji}(t)$ to
reconstruct the sample paths (realizations) of the process.  We
imagine that the process begins in the state $i$.  It waits in that
state for some amount of time given (probabilistically) by
(\ref{eq:waittime}).  When it makes a transition at time $t$, the
probability that it will go to the state $j$ is $\Pi_{ji}(t)$.  Once
in state $j$, it waits for some amount of time, given
(probabilistically) by (\ref{eq:waittime}), and so on.  In this way,
one can reconstruct all of the sample paths of the process, and using
some results from measure theory, one can in fact (under certain
conditions) reconstruct the finite-time transition probabilities from
the knowledge of these sample paths.

However, we will follow a different route, due mainly to Kolmogorov
(1931) and Feller (1940).  To begin, note that the transition
probabilities of a Markov process obey the so-called
Chapman--Kolmogorov equations:
\be
\label{eq:ChapKol}
p_{ji}(t,s) = \sum_{k} p_{jk}(t,s+h) p_{ki}(s+h,s)
\ee
for all $s \leq s+h \leq t$.  The intuitive idea is just that we can
`sum over' an intermediate state of the process.  From the
Chapman--Kolmogorov equations, one can obtain as `limit' equations the
so-called forward and backward Kolmogorov equations.  We state the
theorem here without proof.
\begin{quotation}
\noindent {\bf Theorem 1} (Kolmogorov 1931): If the infinitesimal
parameters, $t_{ji}(t)$, for a Markov process, $V_{t}$, are
well-defined, finite, and continuous for all $t$, then the transition
probability functions, $p_{ji}(t,s)$, for $V_{t}$ are partially
differentiable in $t$ and $s$, and the following hold:
\begin{eqnarray}
\label{eq:forKol}
\frac{\partial}{\partial t}p_{ji}(t,s)   & = &  \sum_k
t_{jk}(t)p_{ki}(t,s), \\[3ex]
\label{eq:backKol}
\frac{\partial}{\partial s}p_{ji}(t,s)  & = &  -\sum_k p_{jk}(t,s)t_{ki}(s).
\end{eqnarray}
\noindent These two equations are called the {\bf forward} and {\bf
backward Kolmogorov equations}, respectively.
\end{quotation}
\noindent Notice that, because by Theorem 1 the $p_{ji}(t,s)$ are
partially differentiable and $p_{ji}(t,t) = \delta_{ji}$, one has
\be
\label{eq:partder}
t_{ji}(t)=\frac{\partial}{\partial t}p_{ji}(t,s)\Big\vert_{s=t}\,.
\ee
Equation (\ref{eq:partder}) justifies our earlier intuitive
interpretation of (\ref{eq:inf1}) and (\ref{eq:inf2}).

Note that Theorem 1 does not yet guarantee that, given a set of
(well-defined, finite, and continuous) infinitesimal parameters,
$t_{ji}(t)$, we can find a Markov process and a set of finite-time
transition probabilities for that process consistent with the
$t_{ji}(t)$.  Theorem 1 already presumes the existence of a Markov
process with finite-time transition probabilities $p_{ji}(t,s)$, and
states two relations (the backwards and forwards equations) between
these probabilities and the infinitesimal parameters.  We must
therefore now ask about the existence of solutions of the Kolmogorov
equations given a set of infinitesimal parameters.  Feller answered
this question by showing how (under certain conditions) to construct a
so-called `minimal' solution to the Kolmogorov equations that is
consistent with the given infinitesimal parameters.  We will review
Feller's construction below, but first we state the theorem.
\begin{quotation}
\noindent {\bf Theorem 2} (Feller 1940):
1. Let $t_{ji}(t)$ be a set of continuous, finite but possibly unbounded
      functions on a finite or infinite open interval $T_1<t<T_2$, satisfying
        \be
        t_{ji} \geq 0,\qquad j\neq i,
        \ee
      and
        \be
        \sum_j t_{ji}(t)=0.
        \ee
      Then there exist absolutely continuous functions $p_{ji}(t,s)$ on
      $T_1<s<t<T_2$ such that
        \begin{eqnarray}
        0\leq p_{ji}(t,s)    & \leq &     1,    \\
        \sum_j p_{ji}(t,s)   & \leq &     1,
        \end{eqnarray}
      and such that the two Kolmogorov equations are satisfied, as well as the
      Chapman--Kolmogorov equations. Further, one has
        \be
        \lim_{\epsilon\rightarrow 0^{+}}p_{ji}(t,t-\epsilon)=\delta_{ji}.
        \ee
2. Under certain conditions, in particular if the index set $I$ is finite,
      one has
        \be
        \sum_j p_{ji}(t,s)=1.
        \ee
For any other solution $q_{ji}(t,s)$ of the Kolmogorov equations satisfying
${\displaystyle \lim_{\epsilon \rightarrow 0^{+}}} q_{ji}(t,t-\epsilon) =
      \delta_{ji}$ one has
        \be
        q_{ji}(t,s)\geq p_{ji}(t,s).
        \ee
      Thus, in this case, the `minimal' solution $p_{ji}(t,s)$
      is the unique solution of the Kolmogorov equations with
${\displaystyle \lim_{\epsilon \rightarrow 0^{+}}} p_{ji}(t,t-\epsilon)
= \delta_{ji}$
      and $\sum_j p_{ji}(t,s) \leq 1$, and hence the only solution that can
      be interpreted as a set of transition functions.
\end{quotation}

We review Feller's construction of the minimal solution here because it
is fairly intuitive.  Feller defines
\be
\label{eq:zerosteps}
p^{(0)}_{ji}(t,s) := \delta_{ji}\exp \left[ { -\int_s^t t_i(u)du }
\right],
\ee
which can be interpreted as the probability that a system in the state 
$i$ at time $s$ will be in the state $j$ at time $t$, having made no 
transitions during the interval $[s,t]$.  (Recall the discussion of 
(\ref{eq:waittime}).) By a careful enumeration of the possibilities, 
one can extend (\ref{eq:zerosteps}) to the probability that a system 
in the state $i$ at time $s$ will be in the state $j$ at time $t$, 
having made $n$ transitions during the interval $[s,t]$:
\be
\label{eq:nsteps}
p^{(n)}_{ji}(t,s) := \sum_k\int_s^t\exp \left[ { -\int_u^t t_j(v)dv}
\right] \Pi_{jk}(u)t_k(u)p^{(n-1)}_{ki}(u,s)du.
\ee
Reading (\ref{eq:nsteps}) from right to left, it is the probability 
that the system starting in state $i$ at time $s$, has arrived at 
state $k$ by time $u$ having made $n-1$ transitions, multiplied by the 
probability that it makes a transition during $[u,u+du]$, multiplied 
by the probability that this transition is to the state $j$, 
multiplied by the probability that it remains in state $j$ until $t$, 
integrated over all intermediate times, $u$, and summed over all 
intermediate states $k$.

Feller defines the finite-time transition functions of the `minimal'
process to be
\be
\label{eq:Fellerdef}
p_{ji}(t,s):=\sum_{n=0}^\infty p^{(n)}_{ji}(t,s).
\ee
He then shows that $p_{ji}(t,s)$ is a solution of both the forward
and the backward Kolmogorov equations with all the stated properties,
including satisfaction of the Chapman--Kolmogorov equations.$^7$

One can further extend Feller's theorem, at least in the case of 
finite $I$, to cover infinitesimal parameters of the form 
$t_{ji}(t)=\Pi_{ji}(t)t_i(t)$ where $t_i(t)$ has singularities, as 
long as these singularities are isolated and integrable 
(Bacciagaluppi, 1996a, Ch.~7, Appendix 2).  However, {\em 
non-integrable} singularities also occur, in fact very naturally, 
namely whenever a probability $p_i(t)$ vanishes.  If $p_i(t)=0$, then 
if $p_i(s)\neq 0$ for some $s<t$, one must also have $p_{ii}(t,s)=0$.  
But now, by (\ref{eq:waittime}), in order for this probability to 
vanish, $t_i$ must have a non-integrable singularity at $t$.  Thus, a 
more specific discussion of existence and uniqueness is needed in the 
general case.

In the next section, we shall show how to derive infinitesimal 
parameters for the evolution of the complete state of a system, as 
defined earlier.  If the corresponding Kolmogorov equations can be 
uniquely solved, this will lead to Markov processes governing the 
evolution of the complete state.  We cannot always prove as yet the 
existence of such a Markov process because of the above difficulties.  
On the other hand, as we shall briefly discuss, in the context of the 
Bohm theory, an analogous problem has been successfully treated 
already, which makes us expect that similar existence results hold 
also in the present case.

\section{Framework for a Modal Dynamics}

\vspace{-30pt}

\subsection{Continuity of the Evolution of Spectral Projections}

It is clear that the set $S(t)$ is in general genuinely
time-dependent.  For example, in the Vermaas--Dieks interpretation, it
is the set of spectral projections of the statistical operator,
$W(t)$.  It is this time-dependence of $S(t)$ that complicates the
discussion of transition probabilities.

Indeed, $S(t)$ is time-dependent in two ways.  First, the {\it
cardinality} of $S(t)$ can change in time. For example, let
\be
\label{eq:easyexample}
W(t) = \cos^{2}(\theta t) P_{1} + \sin^{2}(\theta t) P_{2}
\ee
\noindent for some $\theta$, and some projections, $P_{1} \perp P_{2}$.
For most times, $W(t)$ has two spectral projections, $P_{1}$ and $P_{2}$,
so that for these times $S(t)$ contains two projections.  However,
when $\theta t = n\pi/4$ for any integer $n$, $W(t)$ has just one
spectral projection, so that for these times, $S(t)$ contains just one
projection.

The second kind of time-dependence involves the time-dependence of the
projections in $S(t)$.  Again a simple example is helpful.  Consider a
case where a system evolves freely.  That is, for $W(0) = \sum_i w_i P_i(0)$, 
we have
\be
W(t) = U(t) W(0) U^{\dagger}(t) = \sum_i w_i U(t) P_i(0) U^{\dagger}(t),
\ee
where $U(t)$ is the unitary group generated by the (time-independent)
Hamiltonian of the free system.$^8$ In this case, the cardinality of $S(t)$ 
is constant, but its elements are genuinely time-dependent.

How are we to deal with these forms of time-dependence?  The first
presents a particularly difficult problem, because our aim is to take
as the state space of a stochastic process an index set, $I$, labeling
the elements of $S(t)$.  But, the theory of stochastic processes
generally assumes that the state space is constant.  The
time-dependence of the elements of $S(t)$ presents a slightly less
serious problem.  Suppose that we have solved the first problem, \ie,
that the cardinality of $S(t)$ is indeed constant in time, so that we
have a genuine state space, $I$.  Then we need, for each time, $t$,
some way to associate each element of $S(t)$ with some value in $I$.
Of course, we could simply assume the existence of a family of
bijective maps, $\mu_{t}:I \rightarrow S(t)$, but it would be nice if
instead we could exhibit them.

To begin to resolve these problems, it is useful to discover some
general facts about how the spectral projections of a statistical
operator evolve.  Here we show that, in the finite-dimensional case,
they evolve continuously.  We also give a brief qualitative discussion
of the infinite-dimensional case.  (For a more detailed treatment, see
Bacciagaluppi, Donald and Vermaas (1995).) Continuity makes easy a
resolution of the problems that arise from the genuine time-dependence
of $S(t)$ as it is presently defined (for the atomic version).

We consider a system at a time, $t=0$, with statistical operator
$W(0)$.  $W(0)$ has a spectral resolution
\be
\label{eq:spectral1}
W(0)=\sum_iw_i(0)P_i(0).
\ee
At a time $t$ in a neighborhood of $0$, the system's state will have a
spectral resolution
\be
\label{eq:spectral2}
W(t)=\sum_jw_j(t)P_j(t).
\ee
\noindent We are interested in the relation between these two spectral
resolutions.

Notice that the evolution of $W(t)$ is continuous, \ie, for $t$ 
sufficiently small, $W(t)$ is close to $W(0)$ (in the trace-class 
norm, which is the physically relevant norm for statistical 
operators).  This fact by itself implies nothing for the behavior of 
the spectral projections of $W(t)$ in a neighborhood of $0$.  However, 
it suggests that we consider $W(t)$ as a perturbation of the operator 
$W(0)$ (induced by the time evolution), and apply results from 
perturbation theory in order to compare the eigenprojections of the 
perturbed and unperturbed operators.  In particular, we can use 
Rellich's theorem (see for instance Kato (1966), Rellich (1969), and 
Reed and Simon (1978)), the central theorem of analytic perturbation 
theory.

In the finite-dimensional case, Rellich's theorem states the 
following.  Take a family of operators, $A(z)$, depending on a {\em 
complex} parameter $z$, and such that $A(z)$ is self-adjoint if $z$ is 
real.  If the dependence of $A(z)$ on $z$ is (complex-)analytic in a 
neighborhood of a real $z_0$, then both the eigenvalues and the 
eigenprojections of $A(z)$ are also analytic functions of $z$ in the 
following sense (the assumption cannot be weakened to real 
analyticity, \ie, infinite real differentiability).

If the eigenvalues of $A(z)$ are distinct throughout a neighborhood of
$z_0$, the theorem simply means that the corresponding mutually
orthogonal eigenprojections $P_i(z)$ will be analytic functions of $z$
in a neighborhood of $z_0$.  Instead, if the eigenvalues of $A(z)$
cross in $z_0$, there is a one-to-many correspondence between the
eigenprojections of $A(z)$ at $z = z_0$ and $z\neq z_0$.  However, the
trajectories $P_i(z)$, which are analytic for $z\neq z_0$, can be {\em
analytically extended} also to $z = z_0$.

Thus, if the statistical operators $W(t)$ (extended to a complex
neighborhood of the real line) depend complex-analytically on $t$,
then the eigenprojections $P_i(t)$ will be analytic functions of $t$
with none but {\em removable} singularities, situated at the points in
which the eigenvalues of $W(t)$ cross.  At such points, following
Bacciagaluppi, Donald and Vermaas (1995), we propose to define the
definite properties of the system as the corresponding limits of the
$P_i(t)$.  This proposal is a dynamically motivated {\em extension} of
the modal interpretation.  (See also the discussion by Bacciagaluppi
(1996a, Ch.~6).) To put it differently, we postulate that (for an
atomic system) $S(t)$ is not always given by the spectral resolution
of $W(t)$, but by the set of projections that are the analytic
continuations of the eigenprojections of $W(t)$ when the eigenvalues
cross.

A particularly simple example is afforded by (\ref{eq:easyexample}).
We noted before that in that case, $S(t)$ contains two elements except
when $\theta t = n\pi/4$ for any integer $n$.  However, these
singularities are clearly isolated, and we can postulate that for
times, $t$, such that $\theta t = n\pi/4\ $ for some integer $n$,
$S(t) := \{P_1, P_2\}$.  This modification of the definition of $S(t)$
clearly makes the projections in $S(t)$ evolve analytically.  (Indeed,
they are constant.)

We thus only have to show that if $W(t)$ is a statistical operator, 
then it is an analytic function of $t$.  To do so, it is sufficient to 
consider the so-called `reduced' states, \ie, those obtained by 
partial tracing over the state of some `compound' system.  (We assume 
that the state of the universe is pure, so that its spectral 
resolution obviously evolves continuously.) Let the Hilbert space for 
the subsystem of interest be $\cH^{\ga}$, and let the Hilbert space 
for the rest of the compound system (ultimately, the rest of the 
universe) be $\cH^{\gb}$.  The Hilbert space for the compound system 
is therefore $\cH = \cH^\ga \otimes \cH^\gb$.  We distinguish three 
cases, depending on the dimensionalities of the different Hilbert 
spaces: (1) both $\cH^\ga$ and $\cH^\gb$ are finite-dimensional; (2) 
$\cH^\ga$ is finite-dimensional, but $\cH^\gb$ is 
infinite-dimensional; (3) $\cH^\ga$ is infinite-dimensional.  Here we 
shall treat explicitly only the elementary case (1), with some brief 
remarks on cases (2) and (3).

We assume throughout that the total system is isolated and evolves
according to the Schr\"odinger equation with a constant Hamiltonian,
$H$:
\be
\label{eq:totalevolution}
\ket{\Psi(t)}=e^{-iHt}\ket{\Psi(0)},
\ee
given the initial state $\ket{\Psi(0)}$.  Let $E_1,\ldots,E_N$ and
$\ket{e_1},\ldots,\ket{e_N}$ be the eigenvalues and eigenvectors of
$H$, and write
\be
\label{eq:initialstate}
\ket{\Psi(0)}=\sum_j\lambda_j\ket{e_j}.
\ee
\noindent Then
\be
\ket{\Psi(t)}=\sum_j\lambda_j e^{-iE_jt}\ket{e_j}.
\ee
Introduce any product basis in $\cH = \cH^\ga \otimes \cH^\gb$, with
basis vectors, say, $\ket{\psi^\ga_m}\otimes\ket{\psi^\gb_n}$.  Then,
the reduced state of $\ga$ is given by
\be
W^\ga(t) = \sum_{i,j,k}\lambda_j\overline{\lambda}_k e^{-i(E_j-E_k)t}
\braket{\psi^\gb_i}{e_j}\braket{e_k}{\psi^\gb_i}
\ee
(where the overbar denotes complex conjugation).
Because every $\ket{e_j}$ is some (finite) linear combination
of the $\ket{\psi^\ga_m} \otimes \ket{\psi^\gb_n}$,
\be
\ket{e_j} = \sum_{m,n} c^j_{mn} \ket{\psi^\ga_m} \otimes \ket{\psi^\gb_n},
\ee
\noindent the matrix elements of $W(t)$ will be (finite) linear
combinations of the functions $e^{-i(E_j-E_k)t}$, and thus analytic in
$t$.  This result is what we needed to establish.

For case (2), it can be shown (using estimates on the trace-class norm
of $W^\ga(t)$) that if $\ket{\Psi(t)}$ in (\ref{eq:totalevolution})
depends analytically on $t$, then $W^\ga(t)$ also depends analytically
on $t$.  Analyticity of $\ket{\Psi(t)}$ is assured if one restricts the
initial state (\ref{eq:initialstate}) to a certain dense set in $\cH$.
Finally, in case (3), Rellich's theorem holds also in infinite
dimensions for eigenvalues of the unperturbed operator that are
isolated and of finite multiplicity.  All nonzero eigenvalues of
$W^\ga(t)$ are isolated and of finite multiplicity.  Failure of the
theorem for the eigenvalue $0$ means that trajectories can be born
>from or can die into the null space of $W^\ga(t)$.  A fuller
discussion of cases (2) and (3) is given in Bacciagaluppi, Donald,
and Vermaas (1995).  In the following, we shall assume that $\cH^\ga$ is
finite-dimensional and assume analyticity of $W(t)$.

Using these results, then, the two complications
mentioned earlier are resolved.  In particular, using the extended atomic
version of the modal interpretation, the cardinality of the set $S(t)$
is time-independent, and therefore can be indexed by a fixed index
set, $I$, at all times.  Moreover, continuity makes easy the
definition of the family of functions $\mu_{t}$.  At some initial
time, $t_{0}$, arbitrarily associate with each element of $S(t_{0})$
some element of the index set, $I$.  That is, choose $\mu_{t_{0}}$
arbitrarily.  For all later times, $\mu_{t}$ is determined by the
rule $\mu_t(i) := $ the (continuous) evolute of $\mu_{t_0}(i)$ for
all $i$ and all $t>t_0$.

Using this definition of the maps $\mu_{t}$, we can focus attention on
finding transition probabilities for a stochastic process, $V_{t}$,
whose state space is $I$.  With the help of the map $\mu_{t}$, such a
process induces a stochastic evolution on the elements of $S(t)$ (as
defined by the extended atomic version), which, as we saw, will itself
induce a stochastic evolution of the complete `modal' state of a
physical system (given by the ultrafilters, $m$, as defined in Section
1.1).

As important as these results on continuity are, we note that in {\it
principle}, the problems that we have mentioned can be solved even if
continuity fails.  If the cardinality of $S(t)$ as originally given
changes in time, then find the $S(t)$ with the greatest cardinality
and decompose the elements of $S(t)$ at the other times into
`fiduciary' elements.  For example, suppose that for some $t$, $S(t)$
has the greatest cardinality, $N$, while $S(s)$ (for some $s \neq t$)
has cardinality $N-2$.  In that case, there must be some
multi-dimensional projections in the faux-Boolean algebra generated by
$S(s)$.  Suppose for illustration that $S(s)$ contains a
three-dimensional element, $P$.  Then we can decompose $P$ into three
one-dimensional, mutually orthogonal, projections, $P = P_{1} + P_{2}
+ P_{3}$, replacing $P$ in $S(s)$ with these three projections.  The
resulting faux-Boolean algebra will contain the old one as a
subalgebra.  Hence any dynamics involving the new algebra will induce
a dynamics on the old algebra.  And obviously the maps $\mu_{t}$ can
be defined arbitrarily.

However, we also emphasize that the existence in principle of a
solution to the two problems is far less comforting than the actual
existence provided by the results on analyticity discussed above.
Moreover, as we will see, at least some reasonable dynamics require
that the elements of $S(t)$ be differentiable in time, and without the
results on analyticity, there is no guarantee of differentiability.
In any case, henceforth we ignore these details, and concentrate on
finding $V_{t}$.

\subsection{Infinitesimal Parameters and Single-Time Probabilities}

We must now show how proceeding from a definition of the infinitesimal 
parameters to the finite-time transition probabilities (as in Theorems 
1 and 2), we can ensure that our finite-time transition probabilities 
are consistent with the quantum-mechanical single-time probabilities.  
We do so in this section.

But first:  a remark about notation. Later we will be considering
compound systems, and we will be interested in their single-time joint
probabilities as well as their joint transition probabilities.  The
former we denote by
\[
p_{i,j,\ldots}(t) := p(\ga \ \mbox{has} \ i \
\mbox{at} \ t, \gb \ \mbox{has} \ j \ \mbox{at}\ t, \ldots ).
\]
(Commas separate the indices of different subsystems.)  The latter we
denote by
\[
p_{jn;im}(t,s) := p(\ga \ \mbox{has} \ j \
\mbox{at} \ t, \gb \ \mbox{has} \ n \ \mbox{at}\ t, \ldots \vert
\ga \ \mbox{has} \ i \ \mbox{at} \ s, \gb \ \mbox{has} \ m \ \mbox{at}\
s, \ldots).
\]
(A semicolon separates the later-time indices from the earlier-time
indices.)  However, we continue to denote transition probabilities
for a single system by $p_{ji}(t,s)$.  (These conventions hold also
for the infinitesimal parameters.)

>From the definition of $p_{ji}(t,s)$ and $p_{i}(t)$, it is evident
that
\be
p_{j}(t) = \sum_{i} p_{ji}(t,s) p_{i}(s)
\ee
and
\be
p_{j}(s) = \sum_{i} p_{ij}(t,s) p_{j}(s).
\ee
Therefore, we arrive at the difference equation
\be
\label{eq:difference}
p_{j}(t) - p_{j}(s) = \sum_{i}p_{ji}(t,s) p_{i}(s) - p_{ij}(t,s)
p_{j}(s).
\ee
Using (\ref{eq:difference}), we can write
\be
\label{eq:almostmaster}
p_{j}(t+\epsilon) - p_{j}(t) = \sum_{i}p_{ji}(t+\epsilon,t)
p_{i}(t) - p_{ij}(t+\epsilon,t) p_{j}(t).
\ee
Divide both sides of (\ref{eq:almostmaster}) by $\epsilon$ and take
the limit $\epsilon \rightarrow 0$.  Assuming that the limits exist
(\ie, that the $t_{ji}(t)$ are well-defined), we find that $p_{j}(t)$
is differentiable and that
\be
\label{eq:master}
\dot{p}_{j}(t) = \sum_{i} \left[ { t_{ji}(t) p_{i}(t) - t_{ij}(t)
p_{j}(t) } \right].
\ee
Equation (\ref{eq:master}) is a standard master equation.  (In our
case, it will turn out that $t_{ji}(t)$ is singular only when
$p_{i}(t) = 0$, so that with the convention $\infty \times 0 = 0$,
(\ref{eq:master}) holds even when $t_{ji}(t)$ is singular.) Notice
that we nowhere assumed the Markov property in this derivation of the
master equation.  That is, the $t_{ji}(t)$ could be the infinitesimal
parameters of a non-Markov process.  On the other hand, we have seen
that, given $t_{ji}(t)$, one can construct a canonical Markov process
that has the $t_{ji}(t)$ as infinitesimal parameters, and it is this
route to finite-time transition probabilities that we follow.  (There
can be no construction of a unique non-Markov process from its
infinitesimal parameters because these cannot carry information about
how the {\it history} of the process affects the transition
probabilities.)

The master equation (\ref{eq:master}) is the relation we need between
single-time distributions and infinitesimal parameters.  We have now
to investigate how one can go about solving the master equation for
the $t_{ji}(t)$, {\em given} a set of single-time distributions
$p_j(t)$ and their derivatives.  We shall follow the strategy used by
Bell (1984).  Like Bell, from now on we deal exclusively with the case
of a {\em finite} index set $I$.  (Hence Theorems 1 and 2, with the
above-mentioned limitations, and the results
of Section 3.1 apply.) In fact, our discussion of stochastic processes
clarifies the background and possible limitations also of the work of
Bell (1984), Vink (1993) and Bub (1996, 1997).

Define a {\em probability current} $j_{ji}(t)$ between trajectories,
representing the `net flow' of probability from $i$ to $j$:
  \begin{equation}
    \label{eq:sufficient}
    j_{ji}:= t_{ji}p_i - t_{ij}p_j
  \end{equation}
(where we have suppressed the argument $t$, as we will often do
henceforth).
The definition of $j_{ji}$ implies
  \begin{equation}
    j_{ij} = -j_{ji}.
    \label{eq:condition1}
  \end{equation}
\noindent (That is, the $j_{ji}$ form an antisymmetric matrix.)

Further, equation (\ref{eq:master}) implies a {\em continuity equation}
for the current:
\be
\label{eq:continuity}
\dot{p}_j = \sum_i j_{ji}.
\ee
A solution to (\ref{eq:master}) can thus be obtained by first finding
a current $j_{ji}$ that satisfies (\ref{eq:condition1}) and
(\ref{eq:continuity}), and then, with this $j_{ji}$, finding functions
$t_{ji}$ that satisfy (\ref{eq:sufficient}).  By Theorems 1 and 2, at
least if the $t_{ji}$ are continuous and non-singular,
one can then construct an essentially unique stochastic process
describing the evolution of the system.

Later, we shall exhibit some explicit expressions for $j_{ji}$.  For
now, we consider how to solve (\ref{eq:sufficient}), with the added
constraint that the $t_{ji}$ be a set of (finite-valued) infinitesimal
parameters. That is, we consider how to solve (\ref{eq:sufficient}) under
the constraints
  \begin{equation}
    t_{ji}\geq 0 \qquad(j\neq i),  \label{eq:blah1}
  \end{equation}
and
  \begin{equation}
    \sum_j t_{ji}=0.    \label{eq:blah2}
  \end{equation}
The latter requirement (which is equation (\ref{eq:normalisation})
again) defines $t_{ii}$ in terms of the $t_{ji}$ for $j\neq i$.  In
fact, (\ref{eq:sufficient}) is a system of equations for the $t_{ji}$
with $j\neq i$, the equations being vacuous for $j=i$.  Therefore, we
only need to solve (\ref{eq:sufficient}) under the constraint that the
$t_{ji}$ be non-negative when $i \neq j$.

From (\ref{eq:sufficient}) one has (assuming $p_j>0$)
\be
\label{eq:blah3}
t_{ij}=\frac{t_{ji}p_i-j_{ji}}{p_j}.
\ee
By (\ref{eq:blah1}), $t_{ij}$ must be positive, and therefore we need
$t_{ji}p_i-j_{ji}\geq 0$, or (for $p_i>0$)
\be
t_{ji}\geq\frac{j_{ji}}{p_i},
\ee
whence, again by (\ref{eq:blah1}),
\be
\label{eq:blah4}
t_{ji}\geq\max \left\{ { 0,\frac{j_{ji}}{p_i} } \right\}.
\ee

From Equations (\ref{eq:blah1})--(\ref{eq:blah4}) above, it follows 
that the general solution to (\ref{eq:sufficient}), given $j_{ji}$, 
can be found by choosing for every pair $j<i$ a function $t_{ji}$ that 
satisfies (\ref{eq:blah4}) (and which can be chosen to be continuous).  
The $t_{ij}$ are then uniquely determined by (\ref{eq:blah3}), and the 
$t_{ii}$ by (\ref{eq:blah2}).

The most natural choice for a solution to (\ref{eq:sufficient}) seems 
therefore to be the following: for $j<i$ choose
\be
\label{eq:bell}
t_{ji}:=\max \left\{ { 0,\frac{j_{ji}}{p_i} } \right\}.
\ee
One easily checks with
(\ref{eq:blah3}) that in this case also
\be
t_{ij}=\max \left\{ { 0,\frac{j_{ij}}{p_j} } \right\}.
\ee
Thus, $t_{ji}$ is given by (\ref{eq:bell}) for all $j\neq i$.  In fact,
this is the choice made by Bell (1984) for the solution of
(\ref{eq:sufficient}).

Bell's choice is clearly motivated by the analogy with the guiding 
condition in Bohm's (1952) theory.  Vink (1993) discusses how, in the 
appropriate sense, the Bohm theory is in fact the continuum limit of a 
dynamics of this kind, and how different solutions to 
(\ref{eq:sufficient}) lead to different kinds of theories, one example 
being Nelson's (1985) stochastic mechanics.$^9$

On the other hand, it is also clear from (\ref{eq:sufficient}) that 
whenever $p_i=0$ and $j_{ji}>0$, the infinitesimal parameters are 
singular.  It is actually possible to have $j_{ji}=0$ whenever $p_i=0$ 
(the current we shall construct has this property).  However, as we 
have seen discussing Theorem 2, singularities will arise 
anyway.$^{10}$

We have now a canonical procedure for constructing Markov processes 
with given (differentiable) single-time distributions $p_j$.  We solve 
the linear system of equations (\ref{eq:continuity}) for the $j_{ji}$ 
under the additional constraint (\ref{eq:condition1}).  We then define 
infinitesimal parameters $t_{ji}$ by (\ref{eq:bell}).  If these are 
continuous and regular enough, then by Theorems 1 and 2 we can 
construct a Markov process having the $t_{ji}$ as its infinitesimal 
parameters.  (In the singular case, these theorems are not enough.  
See the remarks in Section 4.3.) One can also see that this Markov 
process will have the given $p_j$ as its single-time distributions, by 
looking at the continuity equation (\ref{eq:continuity}) from the 
inverse perspective, now as a system of linear differential equations 
for the $p_j$ {\em given} the $j_{ji}$.  By the uniqueness theorems 
for systems of ordinary linear differential equations, if two 
stochastic processes (Markovian or non-Markovian) share the same 
infinitesimal parameters and the same initial distribution $p_j(0)$, 
then they will have the same single-time distributions $p_j(t)$ for 
all $t$.

It is clear now that the problem of finding a dynamics, even a 
Markovian one, is vastly {\em underdetermined}.  Even if we limit 
ourselves to Bell's choice (\ref{eq:bell}) in the solution to 
(\ref{eq:sufficient}), the continuity equation (\ref{eq:continuity}) 
has infinitely many solutions that satisfy (\ref{eq:condition1}).  It 
is also clear that the requirement of continuity of the $t_{ji}$ will 
not force one particular choice of current.  Our major remaining task 
is therefore to choose an appropriate solution to the continuity 
equation (\ref{eq:continuity}).

Also this situation has a parallel in the Bohm theory.  Ghirardi and 
Deotto (1997) have recently shown that it is possible to construct 
infinitely many deterministic Bohm-like theories that yield the 
correct single-time position distribution, simply by adding a 
divergence-free term to the standard choice for the probability 
current satisfying the continuity equation
\be
  \frac{d}{dt}\rho(x,t)=\nabla\cdot j(x,t).
\ee
An example of such a Ghirardi--Deotto current can be seen in Bohm and 
Hiley's treatment of the Pauli equation as the non-relativistic limit 
of the Dirac equation (Bohm and Hiley, 1993, section 10.4), where the 
current obtained differs by a divergence-free term from the current 
used in the purely non-relativistic treatment of the Pauli equation 
({\em ibid.}, section 10.2).

\section{An Explicit Dynamics}

\vspace{-30pt}

\subsection{The `Minimal Flow' Current}

We shall be concentrating on a current that we call the `generalized 
Schr\"odinger current', and the dynamics that it induces via 
(\ref{eq:bell}).  But first, we exhibit one other solution, mainly 
because it applies to versions other than the atomic version, if they 
are, indeed, feasible.  (The generalized Schr\"odinger current is 
definable only for the atomic version.)

As we have noted, both (\ref{eq:sufficient}) and (\ref{eq:continuity}) 
are linear systems of equations, to be solved under certain 
constraints.  The case of (\ref{eq:continuity}) is very 
straightforward: the constraints are given by the additional equations 
(\ref{eq:condition1}), which are also linear.  The system consisting 
of (\ref{eq:condition1}) and (\ref{eq:continuity}) has infinitely many 
solutions.  One possible selection criterion would be to minimise the 
overall flow of probability, in the sense that, say,
\be
  \label{eq:minflow}
  \sum_{i,j} j^2_{ji}
\ee
be minimised.  This criterion is mathematically very tractable (in the 
finite case we are now studying), and leads to the following current:
\be
  \label{eq:generalsolution}
  j_{ji} = \frac{1}{D}(\dot{p}_j - \dot{p}_i),
\ee
where $D$ is the number of elements of $S$.  (We omit the proof: see 
Bacciagaluppi (1996a, Ch.~7, Appendix 1).)

Note that this current can be applied whenever we have well-defined 
joint probabilities for the universe, that is, joint probabilities 
covering the joint possession of properties by all subsystems.  For 
example, in the non-atomic version with a preferred factorization 
given by $\cH_{\mbox{\scriptsize univ}} = \cH^{\ga} \otimes
\ldots \otimes \cH^{\gw}$, we would need joint probabilities covering all
atomic systems {\it plus} all combinations of atomic systems.  In a 
universe with only two atomic subsystems, $\ga$ and $\gb$, we could 
write these joint probabilities as
\be
  \label{eq:nonatomic1}
  p_{i_{\ga},i_{\gb},i_{\gab}}(t).
\ee
We would then define the `universal current' to be
\be
  \label{eq:nonatomic2}
  j_{j_{\ga}j_{\gb}j_{\gab};\,i_{\ga}i_{\gb}i_{\gab}} =
  \frac{1}{D}(\dot{p}_{j_{\ga},j_{\gb},j_{\gab}} -
  \dot{p}_{i_{\ga},i_{\gb},i_{\gab}}),
\ee
where $D$ is now the number of elements in the set of all {\it joint} 
states.  Whenever we have differentiable universal joint probabilities 
we can, in principle at least, define a dynamics in this way.

However, such a dynamics has a number of less pleasing properties that 
motivate us to seek a more `quantum-mechanical' current.  For example, 
it is clear that `minimal flow' for the universal current does {\it 
not} entail minimal flow for the currents on subsystems induced by the 
universal current.  Moreover, as we will discuss in detail later, 
there is a standard expression in quantum mechanics for the current 
for a time-{\it independent} set of eigenprojections, and the minimal 
flow current does not reduce to this standard current in the case 
where $S(t)$ happens to be time-independent.

\subsection{Deterministic Evolution for Free Systems}

Before we move on to the task of finding a more satisfactory current, 
we pause to discuss a result first due to Vermaas (1996).  It concerns 
the evolution of the possessed properties of a freely evolving system.  
Under certain fairly natural constraints, Vermaas shows that freely 
evolving systems must evolve deterministically.  This result does 
limit the range of acceptable currents (though there seem to remain 
infinitely many possibilities).  In this section, we give a slightly 
different (and in some ways more limited) justification for the same 
result, mainly because we find the example we shall use to be 
instructive.

We now consider a motivation for the principle that freely evolving 
systems evolve deterministically.  Our motivation is not really a 
`proof', but it is instructive nonetheless.$^{11}$

Consider a system in the initial state
\be
\label{eq:albertsimple}
\ket{\Psi(0)} = \ket{\ga_1}\otimes\ket{\gb_1},
\ee
and let its evolution operator be
\be
  \label{eq:albertev}
  \begin{array}{l}
    U^{\ga} \otimes U^{\gb}= \\[3ex]
    \ \ = \Big( \cos \omega t \ket{\ga_1}\bra{\ga_1} +
    i\sin \omega t \ket{\ga_2}\bra{\ga_1} + \cos \omega t
    \ket{\ga_2}\bra{\ga_2} + i\sin \omega t \ket{\ga_1}\bra{\ga_2} \Big)
    \otimes \ident^{\gb},
  \end{array}
\ee
where $\ident^{\gb}$ is the identity operator on $\cH^{\gb}$.  (This 
operator is unitary.) The system $\ga$ then has its own state vector, 
$\ket{\ga_1}$, so that $\ga$ definitely possesses $P^{\ga}_1$ ($= 
\ket{\ga_1}\bra{\ga_1}$) at time $t = 0$.  The evolution operator 
acting on $\ket{\Psi(0)}$ yields
\be
  \label{eq:albertsimpletime}
  \ket{\Psi(t)} = \left(  \cos \gw t \ket{\ga_1} + i \sin \gw t
  \ket{\ga_2}\right)\otimes\ket{\gb_1},
\ee
so that $\ga$ continues to have its own state vector, $\cos \gw t
\ket{\ga_1} + i \sin \gw t \ket{\ga_2}$, which goes smoothly from
$\ket{\ga_1}$ to $\ket{\ga_2}$ and back again.  Hence $\ga$ `follows' 
this smooth transition.  We might say that in this case the evolution 
of the {\it complete} state of $\ga$ follows the Hamiltonian evolution.  
The same holds if the initial state of the system is instead 
$\ket{\Psi(0)} = \ket{\ga_2}\otimes\ket{\gb_2}$.  In both cases, the 
evolution is necessarily deterministic.

Now, suppose instead that the initial state of the system is
\be
  \label{eq:albertinitial}
  \ket{\Psi(0)} = c_1 \ket{\ga_1}\otimes\ket{\gb_1} +
  c_2 \ket{\ga_2}\otimes\ket{\gb_2},
\ee
but with the same evolution operator, (\ref{eq:albertev}).  In this 
case, $\ga$ initially possesses either $P^{\ga}_1$ or $P^{\ga}_2$.  
Indeed, the reduced statistical operator for $\ga$ is
\begin{eqnarray}
  W^{\ga}(t) & = & |c_1|^2 \Big( \cos^2 \gw t \ket{\ga_1}\bra{\ga_1}
  + i \sin\gw t \cos\gw t \ket{\ga_2}\bra{\ga_1} \nonumber  \\
             &   &
  - i \sin\gw t \cos\gw t \ket{\ga_1}\bra{\ga_2} +
  \sin^2 \gw t \ket{\ga_2}\bra{\ga_2} \Big) \nonumber   \\
             &   &
  + |c_2|^2 \Big( \sin^2 \gw t \ket{\ga_1}\bra{\ga_1} -
  i \sin\gw t \cos\gw t \ket{\ga_2}\bra{\ga_1} \nonumber   \\
             &   & 
  + i \sin\gw t \cos\gw t \ket{\ga_1}\bra{\ga_2} +
  \cos^2 \gw t \ket{\ga_2}\bra{\ga_2} \Big) = \nonumber  \\[1ex]
             & =:& |c_1|^2 {P'}_1(t) + |c_2|^2 {P'}_2(t)\,.
  \label{eq:albertreduced}
\end{eqnarray}
As one would expect, ${P'}_1(t)$ goes smoothly from $P^{\ga}_1$ to
$P^{\ga}_2$ and back again as $t$ goes from $0$ to $\pi/2\gw$, and
conversely for ${P'}_2(t)$.

Knowing that the Hamiltonian takes $\ket{\ga_1}$ smoothly to 
$\ket{\ga_2}$ and vice versa, it is natural to suppose that here too, 
the evolution of the complete state follows the Hamiltonian evolution.  
That is, it is natural to suppose that if the system possesses 
$P^{\ga}_1$ initially, then it will at all times possess ${P'}_1(t)$, 
and similarly for $P^{\ga}_2$ and ${P'}_2(t)$.  Such an evolution is 
deterministic.

This argument can be repeated for any system that evolves freely.  One 
can always ask what the system would do if it were in a pure state 
(corresponding to one of the elements in its spectral resolution, 
ignoring subtleties involving degeneracy, which do not really affect 
the argument).  The answer will be that the complete state {\it must} 
evolve deterministically.  But then it seems reasonable to impose this 
deterministic evolution of the complete state back onto the system 
when its state is not pure.  Hence we shall assume that the complete 
state of any freely evolving system itself evolves deterministically, 
`following' the Hamiltonian evolution.  (We emphasize again that 
Vermaas (1996) {\it has} derived this result from some fairly natural 
conditions.)

\subsection{The Generalized Schr\"odinger Current}

The motivation of the previous subsection was derived partly from a
desire that the possessed properties of systems evolve in a
recognizably `quantum-mechanical' way.  But the minimal flow current
is not at all recognizably `quantum-mechanical'.  Nor does it yield
deterministic evolution for all freely evolving systems.  In this
section, we propose a current that is more quantum-mechanical and
satisfies stability.

As a matter of fact, there is a fairly standard way of deriving a 
probability current from the Schr\"odinger equation.  (See, \eg, 
Cohen-Tannoudji {\it et al.} (1977, pp.~238--240).) This kind of 
current is what is used in the Bohm theory to guide the motion of 
particles.  In Bell's (1984) stochastic modification of the Bohm 
theory, it is used to derive transition probabilities between the 
eigenprojections of the privileged observable (`beable') of the 
theory, supposed to be fermion number density.  Vink (1993) has used 
Bell's treatment to derive a probability current between 
eigenprojections of any given observable, and Bub (1996, 1997) uses 
Bell's and Vink's current, which we call the `Schr\"odinger current', 
to define a dynamics for his own interpretation of quantum theory.

We cannot use the Schr\"odinger current in its usual form, because in 
the atomic version, the privileged observable (defined from 
$S(t)$---see the discussion in Section 1.2) is genuinely 
time-dependent.  However, we shall generalize the Schr\"odinger 
current to the case of a time-dependent privileged observable.  
(Hence, as we said in Section 1.2, our work here is a generalization 
of the work by Bell, Vink and Bub.) We proceed by defining a current 
for general composite systems---and in principle, the universe.  This 
current will give us a dynamics for the joint state of the universe, 
which induces a dynamics for any subsystem.

Recall from Section 1.3 that the state space for the universe is given 
by (\ref{eq:composite}) (and for us, the universe has a 
finite-dimensional Hilbert space given by $\cH_{\mbox{\scriptsize 
univ}} = \cH^\ga \otimes \ldots \otimes \cH^\gw$, so that this state 
space is finite).  The single-time joint probability density is given 
by
\be
  \label{eq:singletime}
  p_{i_\ga, \ldots , i_\gw}(t) :=
  \bra{\psi(t)}
  P^\ga_{i_\ga}(t) \otimes \ldots \otimes P^\gw_{i_\gw}(t)
  \ket{\psi(t)}\,.
\ee
For convenience, we will usually denote the (ordered) set of indices
$(i_\ga, \ldots , i_\gw)$ by a collective index, $i$, so
that (\ref{eq:singletime}) can be written
\be
  \label{eq:easysingletime}
  p_i(t) =
  \bra{\psi(t)} P_i(t) \ket{\psi(t)}.
\ee
Let us assume first that each $P_i$ is one-dimensional and 
time-independent, given by $\ket{r_i}\bra{r_i}$ (so that 
$\{\ket{r_i}\}$ is an orthonormal set, and could be considered the set 
of eigenvectors of the preferred observable, $R$, in Bub's 
interpretation).  Then from the Schr\"odinger equation, it follows 
that
\begin{eqnarray}
  \dot{p}_i(t) & = & -i\braket{\psi(t)}{r_i}\bra{r_i} H \ket{\psi(t)} + 
  i\braket{r_i}{\psi(t)}\bra{\psi(t)} H^{\dagger} \ket{r_i}= \nonumber 
  \\[1ex] 
  \ & = & 2 \Im \Big[ \braket{\psi(t)}{r_i}\bra{r_i}H 
  \ket{\psi(t)} \Big].  \label{eq:pdot}
\end{eqnarray}
We must now choose a current, $j_{ji}(t)$, that satisfies the
continuity equation (\ref{eq:continuity}), where $\dot{p}_j(t)$ is
given by (\ref{eq:pdot}).  The following simple choice is the
`textbook' expression:
\be
  \label{eq:current1}
  j_{ji}(t) := 2 \Im \Big[
  \braket{\psi(t)}{r_j}\bra{r_j}H\ket{r_i}\braket{r_i}{\psi(t)} \Big].
\ee
This current is the one used by Bell and Vink (and, Vink (1993)
argues, in a suitable limit it becomes the current of Bohm's theory).

When the $P_i$ are not one-dimensional, still each $P_i$ is spanned by 
some set of vectors in $\{\ket{r_i}\}$, so that the density, 
time-derivative of the density, and current are obtained by summing 
(\ref{eq:easysingletime}), (\ref{eq:pdot}), and (\ref{eq:current1}) 
over $i$ and $j$ appropriately:
\begin{eqnarray}
  \label{eq:density2}
  p_i(t) & = & \bra{\psi(t)} P_i \ket{\psi(t)}, \\[1ex]
  \label{eq:timeder2}
  \dot{p}_i(t) & = & 2 \Im \Big[ \bra{\psi(t)} P_i H
  \ket{\psi(t)} \Big], \\[1ex]
  \label{eq:current2}
  j_{ji}(t) & = & 2 \Im \Big[ \bra{\psi(t)} P_j H P_i
  \ket{\psi(t)} \Big].
\end{eqnarray}

We now face the problem that, for us, the $P_j$ are genuinely
time-dependent.  Hence in the density in (\ref{eq:easysingletime}),
the $P_i(t)$ are truly time-dependent, and (\ref{eq:pdot}) does {\it
not} give the time-derivative of $p_i(t)$.  Instead, it is
\be
  \label{eq:timeder3}
  \dot{p}_i(t) = 2 \Im \Big[ \bra{\psi(t)} P_i(t) H \ket{\psi(t)} 
  \Big] + \bra{\psi(t)}\dot{P}_i(t)\ket{\psi(t)}.
\ee
We seek a current that satisfies the continuity equation 
(\ref{eq:continuity}) with the left-hand side given by 
(\ref{eq:timeder3}).

Before we go on to find such a current, we note that the approach we
are now following makes two assumptions not required by the `minimal
flow' dynamics of Section 4.1.  First, it assumes that the `universal'
single-time joint probabilities are given by (\ref{eq:easysingletime}).
In a non-atomic version, (\ref{eq:easysingletime}) does not give the
single-time joint probabilities.  Nor is there any obvious way to
represent a universal joint state (one property for each system) by
any projection operator, as there is in the atomic version.  Second,
our present approach assumes that the $P_i(t)$ are differentiable.  By
the results on analyticity that we discussed in Section 3.1, this
assumption is acceptable.

One might wonder why we do not simply stick with the minimal flow 
dynamics, or some other dynamics derived in a similar way, when the 
present approach is so much less general.  The answer is that we find 
the present approach to be much more closely tied to traditional 
methods and questions in quantum mechanics, and while such ties are no 
reason to rule out other approaches, they are sufficient to motivate 
our pursuing this approach.  Indeed, we will now show that there is a 
very natural way to find a generalization of the Schr\"odinger current 
for time-dependent $P_i(t)$.  (The following line of reasoning is due 
to James Cushing (p.~c.).)

For each subsystem, $\gn$, we write $P^\gn_{i_\gn} = 
\ket{\gn_i}\bra{\gn_i}$.  We assume that the universe is in a pure 
state, and we write it in the basis given by the $P^\gn_{i_\gn}$, \ie, 
in the basis $\{\ket{\ga_{i_\ga}} \otimes
\ldots \otimes \ket{\gw_{i_\gw}}\}$:
\be
  \label{eq:universe}
  \ket{\Psi(t)} = \sum_{i_\ga, \ldots , i_\gw} c_{i_\ga, \ldots , 
  i_\gw}(t) \ket{\ga_{i_\ga}(t)} \otimes \ldots \otimes 
  \ket{\gw_{i_\gw}(t)}.
\ee
(If any of the $P^\gn_{i_\gn}$ is more than one-dimensional, then we 
just choose a set, $\{\ket{\gn_k}\}$, such that each $P^\gn_{i_\gn}$ 
is spanned by some subset of $\{\ket{\gn_k}\}$.) As before, we let a 
collective index, $i$, stand in for $i_\ga, \ldots , i_\gw$, so that 
we may write
\be
  \label{eq:simpleuniverse}
  \ket{\Psi(t)} = \sum_i c_i(t) \ket{q_i(t)}
\ee
for some set of (tensor-product) vectors, $\{\ket{q_i(t)}\}$.  
Similarly, we label the $P^\ga_{i_\ga}(t) \otimes \ldots \otimes 
P^\gw_{i_\gw}(t)$ with the collective index $i$.

At every time, $t$, $\{P_i(t)\}$ is a set of mutually orthogonal 
projections, so that there exists some family of unitary operators, 
$\{O(t,s)\}$, such that $P_i(t) = O(t,s) P_i(s) O^{\dagger}(t,s)$ for 
all $t$ and $s$.  Write $O(t):=O(t,0)$.  We define:
\be
  \label{eq:Psiprime}
  \ket{\Psi'(t)} := O^{\dagger}(t)\ket{\Psi(t)},
\ee
so that
\be
  \label{eq:Psiprime2}
  \ket{\Psi'(t)} = \sum_i c_i(t) \ket{q_{i}(0)}.
\ee
Therefore, as far as the atomic version is concerned, $\ket{\Psi'(t)}$ 
differs from $\ket{\Psi(t)}$ only in the fact that its definite-valued 
projections are time-independent (and are, in fact, given by the 
definite-valued projections for $\ket{\Psi(t)}$ at time $t = 0$).  The 
probabilities attached to these time-independent projections are the 
{\it same} as the probabilities attached to their time-dependent 
images under the map given by $O(t)$.  But we already know how to 
write down a current for the time-independent case.  So the obvious 
strategy is to write down the current for $\ket{\Psi'(t)}$, then 
translate the result {\it back} in terms of the time-dependent 
projections, again using the map given by $O(t)$.  In this way, we 
will have derived in a very natural manner a current associated with 
$\ket{\Psi(t)}$, for the time-dependent set $S(t)$.

Using (\ref{eq:Psiprime}) and the Schr\"odinger equation, we have that
\begin{eqnarray}
  {\displaystyle i \frac{\partial \ket{\Psi'(t)}}{\partial t} } & = &
  {\displaystyle
   i \dot{O}^{\dagger}(t) \ket{\Psi(t)} + i O^{\dagger}(t) 
   \frac{\partial \ket{\Psi(t)}}{\partial t}= } \nonumber  \\[1ex]
   \ & = &
  {\displaystyle  i \dot{O}^{\dagger}(t) O(t) \ket{\Psi'(t)} +
  O^{\dagger}(t) H(t) O(t) O^{\dagger}(t) \ket{\Psi(t)}.}
\end{eqnarray}
Defining
\be
  \tilde{H}(t) := i\dot{O}^{\dagger}(t) O(t) + O^{\dagger}(t) 
  H(t) O(t),
  \label{eq:effective}
\ee
we can therefore write
\be
  \label{eq:primedschr}
  i \frac{\partial \ket{\Psi'(t)}}{\partial t} = \tilde{H}(t) 
  \ket{\Psi'(t)}.
\ee
(The operator $\tilde{H}(t)$ is indeed self-adjoint.) As we said, 
$\ket{\Psi'(t)}$ gives rise to time-independent definite-valued 
projections, which are in fact just $P_i(0)$, so that the current 
given by (\ref{eq:current2}) is applicable:
\be
  \label{eq:primedcurrent}
  j_{ji}(t) = 2 \Im \Big[ \bra{\Psi'(t)} P_j(0) \tilde{H}(t) P_i(0)
  \ket{\Psi'(t)} \Big].
\ee
Substituting in for $\tilde{H}(t)$ and using the equality
$O(t)P_j(0)O^{\dagger}(t) = P_j(t)$, we get:
\be
  \label{eq:blahblah}
  j_{ji}(t) = 2 \Im \Big[ i\bra{\Psi(t)} O(t) P_j(0) 
  \dot{O}^{\dagger}(t) P_i(t) \ket{\Psi(t)} + \bra{\Psi(t)} P_j(t) 
  H(t) P_i(t) \ket{\Psi(t)} \Big]
\ee
(where we have also used (\ref{eq:Psiprime})).  Now note that
\begin{eqnarray}
  \dot{P}_j(t) & = & \dot{O}(t) P_j(0) O^{\dagger}(t) + O(t) P_j(0)
  \dot{O}^{\dagger}(t)= \nonumber \\[1ex]
  \ & = &
  \dot{O}(t)O^{\dagger}(t) P_j(t) + P_j(t) O(t) \dot{O}^{\dagger}(t),
  \label{eq:blahblah2}
\end{eqnarray}
so that
\be
  \label{eq:blahblah3}
  O(t)P_j(0) \dot{O}^{\dagger}(t)P_i = \dot{P}_j(t)P_i(t) -
  \dot{O}(t)O^{\dagger}(t)P_j(t)P_i(t).
\ee
But the second term in the right-hand side of (\ref{eq:blahblah3}) is
zero when $i \neq j$ (and when $i = j$ the entire current is zero
anyway, as is clear already from (\ref{eq:primedcurrent})). Therefore
(\ref{eq:blahblah}) is
\begin{eqnarray}
  j_{ji}(t) & = &
  2 \Im \Big[ i\bra{\Psi(t)} \dot{P}_j(t) P_i(t) \ket{\Psi(t)}
  + \bra{\Psi(t)} P_j(t) H(t) P_i(t) \ket{\Psi(t)} \Big]= \nonumber 
  \\[1ex] \ & = &
  2 \Im \Big[ \bra{\Psi(t)} P_j(t) H(t) P_i(t) \ket{\Psi(t)} \Big]
  \nonumber  \\[1ex]
  \ &  &  \ \ \
  + \ \bra{\Psi(t)}\dot{P}_j(t)P_i(t)\ket{\Psi(t)}
  + \bra{\Psi(t)}P_i(t)\dot{P}_j(t)\ket{\Psi(t)}.
  \label{eq:blahblah4}
\end{eqnarray}
Finally, because $P_i(t)P_j(t) = 0$ for $i \neq j$ we have
\be
  \frac{d}{dt} \Big[ P_i(t)P_j(t) \Big]
   = \dot{P}_i(t)P_j(t) + P_i(t)\dot{P}_j(t) = 0,
\ee
so that $P_i(t)\dot{P}_j(t) = - \dot{P}_i(t)P_j(t)$ and the current
in (\ref{eq:blahblah4}) is therefore antisymmetric.  Indeed, we can
rewrite it as:
\be
  \label{eq:genschrcurrent}
  j_{ji}(t) = 2 \Im \Big[ \bra{\Psi(t)} P_j(t) H(t) P_i(t) 
  \ket{\Psi(t)} \Big] + \bra{\Psi(t)}\Big( \dot{P}_j(t)P_i(t) - 
  \dot{P}_i(t)P_j(t) \Big) \ket{\Psi(t)}.
\ee
The current in (\ref{eq:genschrcurrent}) is a natural generalization
of the Schr\"odinger current.  Note in particular that when
$\dot{P}_i(t) = \dot{P}_j(t) = 0$, (\ref{eq:genschrcurrent}) reduces
to the Schr\"odinger current of Bell and Vink.  (Actually, only {\it
one} of $\dot{P}_{i}(t)$ or $\dot{P}_{j}(t)$ need be zero.)

It is also useful to see explicitly that (\ref{eq:genschrcurrent}) 
satisfies the continuity equation (\ref{eq:continuity}).  That it does 
so is clear from the fact that $\sum_i \dot{P}_i(t) = 0$, so that the 
extra term in the generalized current summed over $i$ gives 
$\bra{\psi(t)}\dot{P}_j(t)\ket{\psi(t)}$, which is just the extra term 
in $\dot{p}_j(t)$ as given in (\ref{eq:timeder3}).

We stress, however, that although the current in 
(\ref{eq:genschrcurrent}) is in some sense `natural', it is certainly 
not the only current that generalizes the Schr\"odinger current.  
Indeed, to get a generalized Schr\"odinger current, we need only add 
to the Schr\"odinger current some antisymmetric term that, when summed 
over $i$, gives $\bra{\psi(t)}\dot{P}_j(t)\ket{\psi(t)}$.$^{12}$ For 
example, the expression
\be
  \bra{\psi(t)}\frac{1}{D} \Big( \dot{P}(t) - \dot{P}_i(t) 
  \Big)\ket{\psi(t)},
\ee
which is reminiscent of the minimal flow dynamics, is also an adequate
extra term (where $D$ is the cardinality of $S$.)

In any case, using (\ref{eq:genschrcurrent}) we can now define 
infinitesimal parameters by (\ref{eq:bell}).  By the results of 
Section 3.1, we are allowed to assume analyticity of the probabilities 
$p_i(t)$.  In fact, the $p_i(t)$ will have only isolated zeros, and 
therefore the $j_{ji}(t)/p_i(t)$ will be singular at most on a set of 
isolated points.  Further, it is easy to see also that the $j_{ji}(t)$ 
are analytic, because by the results of Section 3.1 the projections 
$P_i(t)$ are.  Therefore, the singularities of $j_{ji}(t)/p_i(t)$ are 
just poles, and thus both $j_{ji}(t)/p_i(t)$ and 
$t_{ji}(t)=\max\{0,j_{ji}(t)/p_i(t)\}$ are continuous functions with 
values in $[-\infty,\infty]$.

Poles, however, are non-integrable singularities, and thus Theorems 1 
and 2 only ensure that we can canonically construct a Markov process 
for the evolution of the possessed properties {\em between} the 
singularities.  The question of the existence of {\em global} 
solutions is not yet settled.  On the other hand, the situation is not 
surprising and possibly no great cause of concern, if one notes that 
the same problem arises also in the case of the Bohm theory, and that 
further, in the latter case, the existence and uniqueness of global 
solutions has been demonstrated under quite general conditions (Berndl 
{\em et al.}\ 1995).

To be precise, in the Bohm theory the guidance equation,
\be
  \frac{d}{dt}{\bf x}_i(t)
    =\frac{{\bf j}_i({\bf x}_1,\ldots,{\bf x}_N,t)}{\rho({\bf x}_1,
    \ldots,{\bf x}_N,t)},
\ee
will become singular when $\rho({\bf x}_1,\ldots,{\bf x}_N,t)=0$.  
However, Berndl {\em et al.}\ (1995) have shown that for a wide class 
of potentials the guidance equation admits unique global solutions for 
generic initial conditions.  One can expect that for our choice of 
current and for Bell's choice of infinitesimal parameters, results 
similar to Berndl {\em et al.}'s will hold also in the modal 
interpretation.

As a final remark, we show that, as claimed before, $p_i(t) = 0$ 
implies $j_{ji}(t) = 0$.  Writing 
$p_i(t)=\bra{\psi(t)}P_i(t)\ket{\psi(t)}$ it becomes clear that 
$p_i(t) = 0$ if and only if $P_i(t)\ket{\psi(t)} = 0$.  In that case, 
(\ref{eq:genschrcurrent}) is
\be
  \label{eq:jwhenp0}
  j_{ji}(t) = -\bra{\psi(t)}\dot{P}_i(t)P_j(t)\ket{\psi(t)}.
\ee
And, because $P_i(t)\dot{P}_j(t) = - \dot{P}_i(t)P_j(t)$,
(\ref{eq:jwhenp0}) is
\be
  j_{ji}(t) = \bra{\psi(t)}P_i(t)\dot{P}_j(t)\ket{\psi(t)},
\ee
which is zero (under our assumption that $P_i(t)\ket{\psi(t)} = 0$).

In this section, we have discussed the construction of a Markovian 
dynamics at the level of the {\em total} system for the evolution of 
the complete state in the atomic version (at least in the 
finite-dimensional case).  The next section is devoted to the analysis 
of certain aspects of the (generally non-Markovian) dynamics induced 
on the atomic {\em subsystems}.  In particular, we show that the 
choice (\ref{eq:genschrcurrent}) for the current leads to transition 
probabilities whose marginals are deterministic for freely evolving 
atomic systems.

\section{Properties of the Generalized Schr\"odinger Dynamics}

\vspace{-30pt}

\subsection{Determinism for Free Atomic Systems}

Our discussion thus far leaves open the question of the properties of 
the dynamics of subsystems.  Here we will begin to investigate these 
properties.  In this subsection, we will be concerned with the 
dynamics of a freely evolving atomic system.  We will show that such a 
system evolves deterministically, as discussed in Section 4.2.  In the 
next subsection, we make a few remarks towards a general account of 
the dynamics of subsystems.

For simplicity of notation, we concentrate on the atomic system $\ga$, 
and use $i$ and $j$ to label its states.  The joint states of all of 
the other atomic systems in the universe we denote with collective 
indices $m$ and $n$.  To show that when $\ga$ evolves freely, it 
evolves deterministically, we must show that whenever the Hamiltonian 
for the universe takes the form $H = H^{\ga} \otimes 
\ident^{\gb\ldots\gw} + \ident^{\ga} \otimes H^{\gb\ldots\gw}$---\ie, 
whenever there are no interactions between $\ga$ and the rest of the 
universe---the infinitesimal parameters are such that for all $n$ and 
for all $i, m$ such that $p_{im}(t) \neq 0$, $t_{jn;im}(t) = 0$ 
whenever $i \neq j$.

One nice way to prove this claim is by using again the time-dependent 
transformation $O(t)$ used in the derivation of the generalized 
Schr\"odinger current that we gave in the previous section.  Recall 
that there we defined the family of unitary operators, $O(t)$, to be 
such that $P_{i}(t) = O(t) P_{i}(0) O^{\dagger}(t)$, for every 
$P_{i}(t) \in S(t)$.  When the Hamiltonian for the compound system is 
$H = H^{\ga} \otimes \ident^{\gb\ldots\gw} +
\ident^{\ga} \otimes H^{\gb \ldots \gw}$, $\ga$ evolves according to
its {\it own} unitary group,
\begin{equation}
  \label{eq:gaop}
  U(t) = e^{-iH^{\ga}t}.
\end{equation}
Hence the definite-valued projections for $\ga$ evolve according to 
$P^{\ga}_{i}(t) = U(t) P^{\ga}_{i}(0) U^{\dagger}(t)$.  It follows 
that the definite-valued projections for the compound system, which 
may be denoted $P^{\ga}_{i}(t) \otimes P_{m}(t)$, evolve according to
\begin{equation}
  \label{eq:compoundev}
  P^{\ga}_{i}(t) \otimes P_{m}(t) = U(t) \otimes V(t) P^{\ga}_{i}(0)
  \otimes P_{m}(0) U^{\dagger}(t) \otimes V^{\dagger}(t)
\end{equation}
where $\{V(t,s)\}$ (with $V(t):=V(t,0)$) is the appropriate family of 
unitary operators on the rest of the universe.  [Note that $V(t)$ 
bears no simple relation to the evolution operator on the rest of the 
universe, which is $e^{-iH^{\gb\ldots\gw}t}$.]

We are going to substitute $U(t) \otimes V(t)$ for $O(t)$ in 
(\ref{eq:effective}).  To do so, we need to know its derivative as 
well.  It is:
\begin{eqnarray}
  {\displaystyle \frac{d}{dt} \Big[ U(t) \otimes V(t) \Big] }   
  & = &
  {\displaystyle  \frac{d}{dt} \Big[ e^{-iH^{\ga}t} \otimes V(t)
  \Big]= } \nonumber \\[1ex]
  \ & = &  -iH^{\ga}U(t) \otimes V(t) + U(t) \otimes \dot{V}(t).
  \label{eq:derivative}
\end{eqnarray}
Hence the effective Hamiltonian $\tilde{H}(t)$ takes the form
\begin{eqnarray}
  \tilde{H}(t)  & = &
  i\frac{d}{dt}\Big[U(t)\otimes V(t)\Big]^{\dagger} 
  \Big(U(t)\otimes V(t)\Big) + \Big(U(t)\otimes V(t)\Big)^{\dagger} 
  H \Big(U(t)\otimes V(t)\Big) =
  \nonumber   \\
  & = &
  - U^{\dagger}(t)H^{\ga}U(t) \otimes V^{\dagger}(t)V(t)
  + U^{\dagger}(t)U(t) \otimes \dot{V}^{\dagger}(t)V(t)+
  \nonumber    \\
  &   &
  + U^{\dagger}(t)H^{\ga}U(t) \otimes V^{\dagger}(t)V(t)
  + U^{\dagger}(t)U(t) \otimes V^{\dagger}(t)H^{\gb\ldots\gw}V(t)=
  \nonumber    \\
  & = &
  \ident^{\ga} \otimes \Big(\dot{V}^{\dagger}(t)V(t)
  + V^{\dagger}(t)H^{\gb\ldots\gw}V(t)\Big)=
  \nonumber     \\
  & = &
  \ident^{\ga} \otimes \tilde{H}^{\gb\ldots\gw}.
  \label{eq:effective2}
\end{eqnarray}
And thus, (\ref{eq:primedcurrent}) becomes
\be
  j_{jn;im}(t) = 2 \Im \Big[\bra{\Psi'(t)} P^{\ga}_j(0)P^{\ga}_i(0) 
  \otimes  P_n(0)\tilde{H}^{\gb\ldots\gw}P_m(0) \ket{\Psi'(t)}\Big],
  \label{eq:zerocurrent}
\ee
which is obviously zero for $j\neq i$.  And therefore, given the 
choice (\ref{eq:bell}) for the $t_{jn;im}(t)$, we find that 
$t_{jn;im}(t)$ is zero whenever $i \neq j$, providing that $p_{im}(t)
\neq 0$.  Hence we have shown that the freely evolving system, $\ga$,
evolves determinsitically, following the evolution of the
$P^{\ga}_{i}(t)$ according to $U(t)$. (An alternative proof is given by
Bacciagaluppi (1996a, Ch.~7, and 1998a).)

The reader might well be wondering why we did not follow an apparently 
much easier route to the same conclusion.  This easier route would 
have us sum the current, $j_{jn;im}(t)$, over the collective index, 
$m$, to find a `marginal' current for $\ga$.  We could then calculate 
directly the transition probabilities for $\ga$, and it would in fact 
be trivial to show that they are deterministic when $\ga$ evolves 
freely.  But this method is fatally flawed: given Bell's solution 
(\ref{eq:bell}), a little thought shows that one {\it cannot} 
calculate transition probabilities for a subsystem by summing the 
compound system's current over the other indices.

\subsection{General Subsystems}

Hence the calculation of transition probabilities for general 
subsystems of the universe is a non-trivial problem.  We present in 
this section just a summary of minor results on this question.  (For 
further discussion of these and related points about dynamics in modal 
interpretations, see Bacciagaluppi (1996a, 1998a, 1998b) and Dickson 
(1995c, 1998).)

First, the result on deterministic evolution of the previous 
subsection allows one to derive already some (finite-time) transition 
probabilities for special cases of atomic systems that do interact 
with their environment.  This can be done by using a technique due to 
Vermaas (1996).  Namely, transition probabilities $p_{ji}(t,s)$ can be 
derived for an interacting system when either at $t$ or at $s$ its 
definite properties are in one-to-one correlation with the properties 
of a freely evolving system (which we must assume to be atomic).  
Thus, for instance, the transition probabilities for an ideal 
measurement of an arbitrary observable of one of two entangled atomic 
systems can be readily calculated.  Interestingly enough, in this case 
the dependence of the measurement result on the possessed property of 
the measured system has the same form as the Born rule of standard 
quantum mechanics.

Further, one can show (as in Bacciagaluppi (1996a, Ch.~7.4)) that 
deterministic evolution holds true also of an atomic system that is 
interacting with its environment, but whose definite properties {\em 
commute} with the interaction Hamiltonian.  Thus, one can show that an 
ideal measurement of an {\em already possessed} property of an atomic 
system does not disturb the possessed property of the system.  
Similarly, if the system is {\em decohered} by its environment, the 
definite properties of the system will approximately correspond, 
except in cases of strong near-degeneracy, to the eigenspaces of the 
decohering observable (which commutes with the interaction 
Hamiltonian, see \eg, Zurek (1981)).  Hence, an atomic system whose 
only interaction with its environment is by ways of decoherence, will 
also (at least approximately) follow a deterministic evolution.  In 
particular, if one idealises a measuring apparatus as having a 
discrete pointer observable that is decohered (again as in Zurek 
(1981)) and, further, as being an atomic system, then after the 
measurement is completed, the pointer readings of the apparatus will 
not exhibit any stochastic jumps.$^{13}$

The results by Vermaas (1996) and this generalization can be used to 
furnish good examples of the `non-Born-like' form of the transition 
probabilities (unlike the special case mentioned above), and of the 
non-Markovian behavior of interacting subsystems (and at the same time 
indications of when Markovian approximations might be possible).  
Further, they illustrate how, in these toy models, the possessed 
properties of atomic systems play the role of {\em hidden variables}, 
determining in part the outcomes of measurements on the system.  From 
this point of view, the atomic modal interpretation equipped with our 
dynamics has many points of contact with the Bohm theory, not least 
the non-locality resulting from the dependence of the transition 
probabilities of one system on the actually possessed properties of 
other systems (analogously to the guidance equation for one particle 
depending on the positions of the other particles).  Problems arising 
in the Bohm theory can be developed and discussed also in the context 
of the modal interpretation, as frame-dependence of trajectories 
(Dickson and Clifton 1998) and the justification of the distribution 
postulate, \ie, the `initial' single-time probabilities 
(Bacciagaluppi, Barrett and Dickson 1997).

In addition, there are `standard' questions from the theory of 
stochastic processes to be asked.  For example, there are, in fact, 
distinctions to be made among various versions of the Markov property, 
and we may ask which, if any, hold, and under what conditions.  Also, 
there is a well-developed theory of semi-group formulations of Markov 
processes.  It would be worth seeing how to formulate the processes we 
describe here in this way, so that the powerful theory of semigroups 
could be brought to bear on certain questions.

Finally, the work we have done here might be adaptable to other 
interpretations.  In the first place, it might be adaptable to some 
modal interpretation for continuous observables, although such an 
interpretation has yet to be fully worked out (for recent work in this 
direction, see Clifton (1997)).  For this case, there is already in 
place a well-developed theory of diffusion processes (continuous-time 
Markov processes with continuous state spaces) upon which a dynamics 
can draw.  In this case, the powerful theory of stochastic 
differential equations could also be useful, and indeed we may hold 
out the prospect of genuine (generally stochastic) equations of motion 
for the modal interpretation.  And apart >from modal interpretations 
for continuous observables, our work may well apply to other 
interpretations.  For example, it may show how to define a dynamics 
for the quantum logic interpretation (on which, see Dickson (1998)), 
or for the many worlds interpretation.  Of course, these remarks are 
somewhat speculative, but what is clear is that there is more 
interesting work to be done.

{\em Acknowledgements}---We thank David Albert, Jeff Bub, Jeremy 
Butterfield, Rob Clifton, Jim Cushing, Dennis Dieks, Matthew Donald, 
Adrian Kent, Pieter Vermaas, and audiences at the University of 
Minnesota, the University of Oxford and the University of Utrecht for 
their helpful comments.  This paper has been completed during GB's 
tenure of a British Academy Postdoctoral Fellowship.

\newpage
\begin{center}
NOTES
\end{center}
\vspace*{40pt}
\noindent 1. For specific proposals, see Van Fraassen (1979; 1991, Ch.~9),
Kochen (1985), Dieks (1988, 1989, 1994), Healey (1989), Bub (1992, 
1994), and Vermaas and Dieks (1995).

\vspace*{12pt}

\noindent 2.  The reason is clear: we would like to
be able to represent, in a sensible way, probabilities of 
conjunctions, disjunctions, and negations of events, and to do so, we 
need some algebraic operations to represent conjunction, disjunction, 
and negation.  (We mention this point because at least one modal 
interpretation, that of Healey (1989), apparently does {\it not} 
choose a set of possible properties that forms an algebra or partial 
algebra, at least not under the lattice-theoretic operations.)

\vspace*{12pt}

\noindent 3.  For a statement of the {\it problem}, see, for
example, Albert and Loewer (1990, 1993) and Elby (1993).  Discussions 
of and solutions to the original problem were given by Bacciagaluppi 
and Hemmo (1994, 1996), Dickson (1994), Healey (1993a, 1993b), and 
Ruetsche (1995).  Recently, however, serious new problems have 
emerged---see Bacciagaluppi (1996b), Bacciagaluppi, Donald and Vermaas 
(1995), and Donald (1997)---which, we believe, show the inadequacy of 
the non-atomic versions.

\vspace*{12pt}

\noindent 4. There are as well other constraints that might be imposed.
See Clifton (1995a, 1995b, 1996), Dickson (1995a, 1995c), Bub and 
Clifton (1996), Bub (1997), and Dieks (1995) for discussions.  For 
very general discussions of constraints on the algebraic structure of 
the set of definite-valued properties see Bell and Clifton (1995), 
Dickson (1995b, 1996), and Zimba and Clifton (1997).

\vspace*{12pt}

\noindent 5.  A consideration of dynamics also turns out to be
important for the discussion of state preparation followed by 
measurement.  See Bacciagaluppi and Hemmo (1997).  In addition, it has 
been argued elsewhere (Dickson 1995c, 1995d) that one way (though not 
the only way) to make sense of the modal interpretation's denial of 
the projection postulate is to give the {\it un}collapsed state 
dynamical significance (so as not to make it superfluous).  The only 
way to do so convincingly is to exhibit a reasonable dynamics in which 
the uncollapsed state plays a crucial role.  Finally, an account of 
dynamics is crucial in the consideration of Lorentz-invariance---see 
Dickson and Clifton (1997).

\vspace*{12pt}

\noindent 6. See, however, the discussion by Bell (1976) in the context
of the Everett interpretation.

\vspace*{12pt}

\noindent 7.  The definition of $p_{ji}(t,s)$ given in
(\ref{eq:Fellerdef}) also yields an interpretation of the pathological 
case
\be
  \label{eq:pathological}
  \sum_j p_{ji}(t,s)<1.
\ee
In fact, if (\ref{eq:pathological}) holds, one can say there is a 
nonzero probability for an {\em infinite} number of jumps to occur in 
the finite time interval $[s,t]$.  Solutions with 
(\ref{eq:pathological}) are called {\em quasi-processes}, or {\em 
dishonest}, or {\em non-conservative} processes.  For example, if $j$ 
represents the number of individuals in a population, $\sum_j 
p_{ji}(t,s)<1$ means that there is nonzero probability for a 
transition from $i$ individuals to infinitely many individuals in the 
finite time interval $[s,t]$.

On the other hand, Feller shows that if the $t_{ii}(t)$ are
bounded uniformly in $i$ by a function $\pi(t)$ such that
\be
  \int_{T_1}^{T_2}\pi(t)^{\alpha}dt<\infty
\ee
for some $\alpha>1$, then
\be
  \sum_j p_{ji}(t,s)=1.
  \label{eq:sumtoone}
\ee
In particular, if $I$ is finite, then ${\displaystyle 
\pi(t):=\max_i\abs{t_{ii}}}$ is integrable to any power $\alpha>1$ on 
every open interval $]s,t[$ with $T_1<s<t<T_2$.  Thus, 
(\ref{eq:sumtoone}) is satisfied on any such interval, and 
consequently on the whole of $]T_1,T_2[$.

\vspace*{12pt}

\noindent 8.  A clarification of `free evolution':  it might be 
supposed that `unitary evolution' and `free evolution' are equivalent 
in quantum mechanics.  They are not.  A system evolves freely between 
times $s$ and $t$ if and only if each member of the family of unitary 
operators $U(t,t')\ (s \leq t' \leq t)$ that carries the system's 
state from time $t'$ to time $t$ is a homogenous function of time,
\ie, is a function only of the difference $t-t'$, in which case
$U(t,t')$ is generated by some time-independent Hamiltonian.

\vspace*{12pt}

\noindent 9.  Actually, Bell's expression for $t_{ji}$ is not exactly
(\ref{eq:bell}), but
\be
  \label{eq:reallybell}
  t_{ji} := \left\{
  \begin{array}{ll}
    {\displaystyle \frac{j_{ji}}{p_i}} \ \ & \ \ \mbox{for} 
    \ j_{ji} > 0, \\[2ex]
     & \ \ \mbox{for} \ j_{ji} \leq 0.
  \end{array} \right.
\ee
The difference arises precisely when $j_{ji}=p_i=0$, where 
(\ref{eq:bell}) may be infinite, while (\ref{eq:reallybell}) is zero.  
Our choice is continuous at these exceptional points.

\vspace*{12pt}

\noindent 10. Here is a direct proof. Suppose that for all $j$,
  \begin{equation}
    \int_{T_0-\epsilon}^{T_0}\abs{\frac{j_{ji}(t)}{p_i(t)}}dt<\infty,
    \label{eq:finite}
  \end{equation}
where $T_0$ is a zero of $p_i(t)$. Then also
\begin{eqnarray}
  \lefteqn{{\displaystyle
  \abs{\int_{T_0-\epsilon}^{T_0}
  \frac{\sum_j j_{ij}(t)}{p_i(t)}dt}=}}
  \nonumber \\[2ex]   
  & & {\displaystyle
  =\abs{\int_{T_0-\epsilon}^{T_0}\frac{\sum_j j_{ji}(t)}{p_i(t)}dt} 
  \leq
  \sum_j\int_{T_0-\epsilon}^{T_0}\abs{\frac{j_{ji}(t)}{p_i(t)}}dt 
  <\infty, }
\end{eqnarray}
by (\ref{eq:condition1}), the triangle inequality and 
(\ref{eq:finite}).  However, $\sum_j j_{ij}(t)=\dot{p}_i(t)$, by 
(\ref{eq:continuity}), and $\dot{p}_i(t)/p_i(t)$ has a logarithmic, 
thus non-integrable, singularity at $T_0$:
\begin{eqnarray}
  \lefteqn{{\displaystyle
  \int_{T_0-\epsilon}^{T_0}\frac{\dot{p}_i(t)}{p_i(t)}dt=}}
  \nonumber \\[2ex]
  & & {\displaystyle
  =-\int_{T_0-\epsilon}^{T_0}\frac{d}{dt}\log p_i(t)dt=
  -\log p_i(t)\Big|_{T_0-\epsilon}^{T_0}=\infty. }
\end{eqnarray}
Thus, at least for some $j$, (\ref{eq:finite}) must fail.

\vspace*{12pt}

\noindent 11.  The example to follow is based on an example suggested to
us by David Albert at the Workshop on Quantum Measurement, University 
of Minnesota, May 1995.

\vspace*{12pt}

\noindent 12. From this point of view, we can see the construction of the
current in (\ref{eq:genschrcurrent}) as follows.  Two expressions 
that, when summed over $i$, obviously yield the extra term in 
$\dot{p}_j(t)$ are
\be
  \label{eq:term1}
  \bra{\psi(t)} \dot{P}_j(t) P_i(t) \ket{\psi(t)}
\ee
and
\be
  \label{eq:term2}
  \bra{\psi(t)} P_i(t) \dot{P}_j(t) \ket{\psi(t)}.
\ee
\noindent Neither (\ref{eq:term1}) nor (\ref{eq:term2}) are in general real,
but they are complex conjugates of one another.  Hence by adding them, 
we get a real number, and (using $P_i(t)\dot{P}_j(t) = - \dot{P}_i(t) 
P_j(t)$) their sum is anti-symmetric and, indeed, just the term that 
we added to the Schr\"odinger current to get 
(\ref{eq:genschrcurrent}).

\vspace*{12pt}

\noindent 13. The requirements of decoherence (and thus macroscopicity) of
the pointer and at the same time of atomic behavior (in the sense of 
our dynamics) will not necessarily go hand in hand.  Further, as 
hinted in note 3, in more realistic (infinite- or high-dimensional) 
models, not even the definiteness of pointer readings seems given (see 
in particular Bacciagaluppi (1996b)).  As toy models for the dynamics, 
however, these examples serve their purpose.

\end{document}